\documentclass[12pt,a4paper]{article}

\usepackage[english]{babel}
\usepackage[autostyle,english=british]{csquotes} 

\usepackage[
  backend=biber,
  bibstyle=ieee,
  citestyle=numeric-comp
]{biblatex}
\usepackage{mathtools,amssymb,bbm}
\usepackage{braket,tensor}
\usepackage{xcolor}
\usepackage{textgreek}
\usepackage{nicefrac}
\usepackage{amssymb}
\usepackage{amsfonts}

\usepackage{enumitem}
\usepackage[a4paper, total={6in, 9in}]{geometry}
\usepackage[hidelinks]{hyperref}
\definecolor{orange}{RGB}{255, 77, 0}
\hypersetup{
 colorlinks,
 linkcolor={orange},
 citecolor={orange},
 urlcolor={orange}
}

 \csname
@addtoreset\endcsname{equation}{section}

\newcommand{\diff}{\mathrm{d}}
\newcommand\must{\mathrel{\stackrel{\makebox[0pt]{\mbox{\normalfont \small !}}}{=}}}

\newcommand{\be}{\begin{equation}}
\newcommand{\ee}{\end{equation}}

\newcommand{\ba}{\begin{equation} \begin{aligned}}
\newcommand{\ea}{\end{aligned} \end{equation}}

\newcommand{\bea}{\begin{eqnarray}}
\newcommand{\eea}{\end{eqnarray}}

\newcommand{\bpsi}{\bar{\psi}}
\newcommand{\pa}{\partial}
\newcommand{\balpha}{\bar{\alpha}}

\newcommand{\bL}{\bar{L}}
\newcommand{\bbeta}{\bar{\beta}}
\newcommand{\btheta}{\bar{\theta}}

\newcommand{\bQ}{\bar{Q}}

\newcommand{\bC}{\bar{\mathcal{C}}}
\newcommand{\bB}{\bar{\mathcal{B}}}
\newcommand{\C}{\mathcal{C}}
\newcommand{\B}{\mathcal{B}}
\newcommand{\Q}{\mathcal{Q}}
\newcommand{\J}{\mathcal{J}}

\newcommand\smallO{
 \mathchoice
 {{\scriptstyle\mathcal{O}}}
 {{\scriptstyle\mathcal{O}}}
 {{\scriptscriptstyle\mathcal{O}}}
 {\scalebox{.7}{$\scriptscriptstyle\mathcal{O}$}}
 }

\begin{document}

\begin{titlepage} 
\begin{center}
{\LARGE \bf Pair production of massive charged\\[1mm]vector bosons from the worldline}
\vskip 1.2cm
Fiorenzo Bastianelli$^{\,a,b}$, Filippo Fecit$^{\,a,b}$ and Alessandro Miccich\`e$^{\,a}$
\vskip 1cm
$^a${\em Dipartimento di Fisica e Astronomia ``Augusto Righi", Universit{\`a} di Bologna,\\
via Irnerio 46, I-40126 Bologna, Italy}\\[2mm]
$^b${\em INFN, Sezione di Bologna, via Irnerio 46, I-40126 Bologna, Italy}\\[2mm]
\end{center}
\vskip 1cm

\abstract{
We investigate a worldline formulation for a massive spin-1 particle interacting with an electromagnetic background. Two first-quantized descriptions of the spin degrees of freedom are considered: one based on bosonic oscillators and the other on fermionic oscillators. Focusing initially on the bosonic model -- which can accommodate particles of arbitrary integer spin -- we review how quantization in the spin-1 sector, performed both via Dirac's method and BRST quantization, reproduces the free Proca field theory.

We then introduce coupling to an external electromagnetic field and demonstrate that Maxwell's equations for the background emerge as a consistency condition for the nilpotency of the BRST charge on the spin-1 sector.
Encouraged by this result, which proves the viability of the particle model, we proceed to construct a path integral quantization of the worldline action for the charged spin-1 particle on the circle. This yields the one-loop effective Lagrangian for a constant electromagnetic field induced by a massive charged vector boson. As expected, the result reveals a vacuum instability, which we quantify by deriving the pair production rate for the vector bosons, recovering
previous results obtained in quantum field theory.

For comparison, we repeat the analysis using the standard $\mathcal{N}=2$ spinning particle model, 
which contains fermionic worldline degrees of freedom, finding identical results. 

Finally, we comment on possible extensions of the worldline models to include effective interactions and briefly explore their implications for pair production.}

\end{titlepage}

\tableofcontents

\section{Introduction}
Although experimental evidence for non-perturbative particle-antiparticle pair production remains elusive, theoretical investigations have made substantial progress in uncovering its underlying mechanisms. The pioneering work of Julian Schwinger \cite{Schwinger:1951nm} in the 1950s demonstrated that a constant, strong electric field induces vacuum instability in a fermionic quantum field theory, leading to the creation of electron-positron pairs. Building up on this simple scenario, the need of a better understanding of the phenomenon has prompted extensive studies of more realistic and rich situations, such as the effect of spatially inhomogeneous electromagnetic fields \cite{Dunne:2005sx} and the possibility of enhancing the process with the presence of an additional rapidly oscillating field, leading to a dynamically assisted pair production \cite{Schutzhold:2008pz}.

In this context, the functional approach known as the Worldline Formalism \cite{Strassler:1992zr, Schubert:2001he, Newbook:2025} has emerged as a particularly powerful framework, offering a natural setup to perform non-perturbative analyses. This formalism has been extensively applied as an alternative to conventional second-quantized methods, to both scalar and spinor Quantum Electrodynamics (QED), as reviewed in \cite{Schubert:2001he} and further expanded in \cite{Ahmadiniaz:2020wlm, Ahmadiniaz:2021gsd}, with deeper insight enabled by the development of worldline instanton techniques \cite{Affleck:1981bma, Dunne:2005sx, Dunne:2006st, Amat:2022uxq, DegliEsposti:2022yqw}. The Worldline Formalism has then been adapted to cover pair creation in other interesting situations; these include the study of the vacuum instability of other quantum fields in different classical backgrounds, e.g. the case of a quantum scalar field coupled to a Yukawa background \cite{Fecit:2025kqb}, which may be readily extended to include axial interactions building on the results of \cite{Bastianelli:2024vkp}, and the dual Schwinger effect for production of magnetic monopoles in strong magnetic fields \cite{Gould:2021bre}. These approaches typically adopt a ``top-down" strategy, starting from quantum field theory expressions, most often the diagonal of the heat kernel which is related to the vacuum persistence amplitude, and reformulating them in terms of a suitable worldline representation.

In this work, we adopt a more principled, ``bottom-up", approach, without relying on any pre-derived quantum field theory (QFT) input. We aim to study a charged, massive spin-1 particle and compute the one-loop effective Lagrangian it induces in a constant electromagnetic (EM) background. This effective Lagrangian encodes information about vacuum instability and pair production.
Historically, a similar Lagrangian was first derived by Euler and Heisenberg by studying electron loops \cite{Heisenberg:1936nmg}, and was shortly thereafter extended by Weisskopf to include massive charged scalars  \cite{Weisskopf:1936hya}.
The corresponding result for massive spin-1 charged particles was obtained much later \cite{Vanyashin:1965ple}, and our goal here
is to reproduce it using genuine worldline methods. A review of Euler and Heisenberg type Lagrangians is available in \cite{Dunne:2004nc, Dunne:2012vv}, while their derivation via worldline techniques can be found in \cite{Schubert:2001he}.
The spin-1 case, which we wish to address here, presents several subtleties. In particular, it is well-known that massless charged spin-1 particles are inconsistent due to the breaking of their own gauge invariance by the electromagnetic coupling \cite{Weinberg:1980kq}. 
In contrast, massive charged spin-1 particles are theoretically consistent, as exemplified in the Standard Model, where 
the $W^\pm$ bosons happily interact with electromagnetism, and thus one should be able to describe them consistently within a worldline approach.

To proceed, we must construct a worldline action capable of describing a massive charged spin-1 particle, hence reproducing the Proca theory \cite{Proca:1936fbw}. 
Traditionally, worldline models have been used to describe spin $s$ particles relying on fermionic variables to parametrize the spin degrees of freedom and local $\mathcal{N}=2s$ worldline supersymmetries to ensure unitarity \cite{Berezin:1976eg, Barducci:1976qu, Brink:1976sz, Gershun:1979fb, Howe:1989vn}. 
However, the couplings to background fields for $s \geq 1$
 may break worldline supersymmetry (SUSY), and this fact often leads to inconsistencies. 
 In this regard, BRST methods have proven particularly useful for determining consistent couplings.
 They have been successfully employed in several recent works \cite{Dai:2008bh, Bonezzi:2018box, Bonezzi:2020jjq, Carosi:2021wbi, Fecit:2023kah}, establishing them as one of the most effective frameworks for studying the conditions required to achieve consistent couplings in worldline models. We are going to use this approach.

To identify the action, we begin by choosing a bosonic, rather than fermionic, worldline model: we take a 
spinning particle with bosonic oscillators and 
``bosonic supersymmetry" on the worldline \cite{Henneaux:1987cp, Barnich:2004cr, Hallowell:2007qk, Bastianelli:2009eh} and build on the recent BRST analysis of \cite{Bonezzi:2024emt}. Such bosonic oscillators have also been recently employed 
in \cite{Haddad:2024ebn, Hoogeveen:2025tew} for studying black hole scattering using worldline methods.
The model analyzed in \cite{Bonezzi:2024emt} is seen to contain massless bosonic excitations of any spin. 
Therefore, we first extend it to accommodate a mass term and then focus on the spin $s=0,1$ sectors, 
which are expected to admit a consistent coupling to an external electromagnetic field. 
 For spin 0, we find through a BRST analysis that there are no restrictions on the background.
For spin 1, we discover that the nilpotency of the BRST charge requires the background to satisfy Maxwell's equations. 
This is enough to prove the consistency of the electromagnetic coupling,
so that we may turn with confidence to develop a path integral formulation to compute the one-loop effective action in a constant electromagnetic background, and specifically for spin 1. 
As expected, an imaginary part emerges, signaling potential pair production of massive vector bosons. This way, we reproduce from a first-quantized perspective the findings originally due to Vanyashin and Terent’ev \cite{Vanyashin:1965ple}.
As for the higher spin sectors of the model, the BRST analysis shows that the electromagnetic background must vanish, so that no coupling is possible. 
As anticipated, the spin-0 sector also admits an electromagnetic coupling, and we study this case within the same model with bosonic oscillators, 
with a projection to spin 0 rather than spin 1,
to provide a comparative scenario and as a testbed for our methods. As expected, we recover the one-loop effective Lagrangian for a scalar particle originally due to Weisskopf \cite{Weisskopf:1936hya}.

A spinning particle model with fermionic oscillators can also be employed. The standard
$\mathcal{N}=2$ spinning particle, which features two local supersymmetries on the worldline, is capable of describing spin 1 and/or antisymmetric tensor fields \cite{Howe:1989vn}. 
Coupling this model to a gravitational background does not present significant challenges: the defining $\mathcal{N}=2$  
first-class constraint algebra remains first-class even in curved space, ensuring that local supersymmetry is preserved. The coupled model is consistent and has been considered in \cite{Bastianelli:2005vk, Bastianelli:2005uy}
 to analyze the gravitational effective action induced by a loop of spin-1 and antisymmetric tensor fields ($p$-forms), both in the massless and massive cases.
Coupling to electromagnetism, however, turns out to be more subtle. The analysis in \cite{Howe:1989vn} suggests that such a coupling breaks supersymmetry. Nonetheless, we consider a massive extension of the model and show, again using BRST techniques, that a coupling to electromagnetism for massive spin-1 excitations is actually feasible. The interaction appears in a manner quite analogous to the treatment of the particle with bosonic oscillators.
We use this alternative worldline model for charged, massive spin-1 particles and find that it yields results identical to those obtained from the previous model based on bosonic oscillators. 

We conclude by commenting on possible extensions of the particle models to include additional effective couplings to electromagnetism, which are expected to take into account the potential non-point-like nature of the particle in question.

We find our results highly satisfying. They provide a test of the intrinsic self-consistency of the worldline ``bottom-up" approach.
We believe our methods and results may be applied to more general analyses: for instance, 
one could extend them to go 
beyond the constant-field case with relatively minor adjustments. Additionally, one could study scattering amplitudes 
within this formalism along the lines pioneered  in \cite{Dai:2008bh} and studied more recently in \cite{Bastianelli:2025xx}.
\\

The paper is organized as follows. In Section \ref{sec1}, we introduce the bosonic worldline model and analyze its free spectrum, mostly focusing on the spin-1 sector. In Section \ref{sec2}, we couple the model to a classical abelian background via BRST quantization, showing that quantum consistency requires the background field to be on-shell. In Section \ref{sec3}, we derive the worldline representation of the one-loop effective action in a constant electromagnetic background, extract its imaginary part, and discuss the implications for Schwinger-type pair production of massive spin-1 particles. We relegate to Section \ref{sec4} a discussion of an alternative formulation based on the fermionic spinning particle with $\mathcal{N}=2$ worldline supersymmetries, reproducing the same results obtained via the bosonic model. Finally,  in Section \ref{sec5}, we consider extensions of the worldline model to include effective interactions and briefly explore their consequences for pair production. Section \ref{sec6} contains our conclusions and outlook.
Technical computations of functional determinants are presented in Appendix \ref{appA}.

\section{Free bosonic spinning particle}\label{sec1}
The worldline model known as ``bosonic spinning particle" is defined by the usual set of phase-space variables $(x^\mu, p_\mu)$, augmented by an additional pair of complex bosonic variables $(\alpha^\mu, \balpha^\mu)$. The former represent cartesian coordinates and conjugate momenta of a relativistic particle, while the latter are needed to account for the spin degrees of freedom. Both pairs are functions of the proper time, which we take to range as $\tau\in[0,1]$. The kinetic term
\begin{equation}
 S_{\mathrm{kin}}=\int \diff\tau \left( p_\mu \dot{x}^\mu -i \bar\alpha_\mu \dot{\alpha}^\mu\right)
\end{equation}
defines the phase space symplectic structure and fixes the Poisson brackets to 
\begin{equation}
\{x^\mu, p_\nu\}_{\mathrm{PB}} = \delta^\mu_\nu 
\;, \quad 
\{\alpha^\mu, \bar \alpha^\nu\}_{\mathrm{PB}} = i \eta^{\mu\nu}
\;.
\end{equation}
As it stands, the model is not unitary, as upon quantization negative norm states will be generated by the $(x^0, p^0, \alpha^0, \balpha^0)$ variables. Moreover, the model, as we shall discuss, contains particle excitations of any integer spin, and one needs to eliminate some further degrees of freedom to describe a single particle with fixed spin. Both problems can be addressed by gauging suitable constraints: the gauged worldline action 
we are interested in is given by
\begin{equation} \label{massless bosonic action}
S=\int \diff\tau\; \left[ p_\mu \dot{x}^\mu -i \bar\alpha_\mu \dot{\alpha}^\mu-eH - \bar{u} L - u \bar{L} - a J
\right]\;, 
\end{equation}
where we introduced the worldline gauge multiplet $(e,\bar u, u, a)$ acting as a set of Lagrange multipliers 
that enforce the constraints
\begin{gather}\label{constraints}
 H = \frac{1}{2}p^\mu p_\mu \;, \quad L = \alpha ^\mu p_\mu\;, \quad \bar{L} = \bar{\alpha}^\mu p_\mu\;, \quad J = \alpha^\mu \bar{\alpha}_\mu \;.
\end{gather}
The latter satisfy a first-class Poisson-bracket algebra:
\begin{equation} \label{algebra}
\{L,\bar L\}_{\mathrm{PB}}= 2i H \;, \quad \{J, L\}_{\mathrm{PB}}= -i L \;, \quad \{J,\bar L\}_{\mathrm{PB}}= i \bar L \;.
\end{equation}
The phase-space functions \eqref{constraints} play rather different roles.
\begin{itemize}
 \item The role of $(H, L, \bL)$ constraints is to remove the negative-norm states and \emph{must} be gauged to make the model consistent with unitarity. The Hamiltonian constraint $H$ corresponds to the mass-shell condition for massless particles, and generates $\tau$-reparametrization in phase space, while the remaining pair, $L$ and  $\bL$, generates ``bosonic" supersymmetries.\footnote{We deliberately use this terminology since \eqref{algebra} is formally similar to a SUSY algebra, except for the fact that $L$ and $\bar L$ are bosonic rather than fermionic.}
\item The $J$ constraint is a $U(1)$ generator which rotates the bosonic oscillators by a phase; its gauging is optional as far as unitarity is concerned. However, upon quantization, it projects the  Hilbert space onto the physical 
subspace with a specific occupation number, describing the degrees of freedom of a particle with maximal spin $s$. For this to happen, one must add a Chern-Simons term on the worldline
with the Chern-Simons coupling fine-tuned according to the value of the spin $s$ one wants to achieve.
This approach has been discussed extensively 
 in \cite{Bastianelli:2013pta, Bastianelli:2015iba}. See also
 \cite{Bastianelli:2021rbt} for a related application to the wordline description of a bi-adjoint scalar. 
  \end{itemize}

To make explicit the gauge symmetries enjoyed by \eqref{massless bosonic action}, one has to compute the action of the constraints on generic phase-space functions $F$ via the Poisson brackets: $\delta F = \{ F, V\}_{\mathrm{PB}}$. Considering the linear combination of the constraints $V=\epsilon H + \bar \xi L + \xi \bar L +\phi J$ with gauge parameters $(\epsilon, \bar \xi, \xi, \phi)$, the corresponding transformations are
\begin{subequations}\label{gauge trans0}
\begin{align}
\delta x^\mu &= \epsilon p^\mu + \xi \bar{\alpha}^\mu + \bar{\xi} \alpha^\mu\;, \label{time-transl}\\
\delta p_\mu &= 0\;, \\
\delta \alpha^\mu &= i\xi p^\mu + i \phi \alpha^\mu\;, \label{gauge1} \\
\delta \bar{\alpha}^\mu &= -i\bar{\xi} p^\mu - i \phi \bar{\alpha}^\mu\;. \label{gauge2}
\end{align}
\end{subequations}
For the action \eqref{massless bosonic action} to be invariant, the gauge fields must transform as follows
\begin{subequations}
\begin{align}
\delta e &= \dot{\epsilon} + 2i u \bar{\xi} - 2i \bar{u} \xi\;, \\
\delta u &= \dot{\xi} - i a \xi + i \phi u\;,  \label{u}
\\
\delta \bar{u} &= \dot{\bar{\xi}} + i a \bar{\xi} - i \phi \bar{u}\;, 
\label{bar u}
\\
\delta a &= \dot{\phi}\;.
\end{align}
\end{subequations}
The need for the worldline constraints to enforce unitarity remains somewhat obscure in this setup. 
To review and clarify this point, it may be beneficial to perform a brief lightcone analysis.

\subsection{Lightcone analysis}
Despite the loss of manifest covariance, a lightcone analysis allows for a direct calculation of the number of propagating physical degrees of freedom. It is a well-known method, employed in many worldline models, see e.g. \cite{Siegel:1988yz, Bastianelli:2014lia, Bastianelli:2015tha}. We define lightcone coordinates $x^{\pm}$ in $D$ spacetime dimensions by
\begin{equation}
 x^\mu=(x^+,x^-,x^a)\;, \quad \text{with} \quad x^{\pm}=\frac{1}{\sqrt{2}}(x^0 \pm x^{D-1})\;,
\end{equation}
where $x^{a=1,\dots,D-2}$ are the transverse directions. The line element reads $\diff s^2=-2 \diff x^+ \diff x^-+\diff x^a \diff x^a$, whence, for any vector $V^\mu$, $V^+=-V_-$ and $V^-=-V_+$.

The guiding idea behind the lightcone analysis is to remove negative-norm states by implementing a gauge-fixing that isolates the physical degrees of freedom, which in turn lead to a manifestly positive-norm Hilbert space upon quantization. To do that, let us first assume motion with $p^+\neq0$ and consider the Hamiltonian constraint 
\be
H= \frac{1}{2}p^\mu p_\mu= -p^+p^- + \frac12 p^ap^a=0 \;.
\ee
It generates time translations, see \eqref{time-transl}. These symmetries   
are gauge-fixed by imposing the lightcone gauge 
\be
x^+=\tau \;.
\ee
Correspondingly, the Hamiltonian constraint is solved for the momentum $p^-$, conjugate to $x^+$,
\be
p^-  = \frac{1}{2p^+} p^ap^a \;.
\ee
At this point, the remaining independent phase-space variables are $(x^-, p^+)$ and $(x^a,p^a)$. The lightcone gauge has the flavor of nonrelativistic mechanics, but the model is fully relativistic. A Hilbert space can be constructed by quantizing these independent variables to obtain a positive-definite Hilbert space. 

On top of these variables, there are also the relativistic oscillators, which may also lead to negative norms.
That this does not happen (the so-called no-ghost theorem) is again made explicit by completing the lightcone gauge
fixing.
The gauge symmetries generated by $L$ and $\bar L$, see Eqs. 
\eqref{gauge1} and \eqref{gauge2}, are fixed by setting 
\begin{equation}
 \alpha^+=0\;, \quad \bar\alpha^+=0\;,
 \label{lcgf}
\end{equation}
while the constraints $L=\bar L=0$ are solved explicitly by expressing the 
variables conjugate to \eqref{lcgf} in terms of the remaining independent variables
\begin{align}
  \bar{\alpha}^-=\frac{1}{p^+}\bar\alpha^a p_a
  \;, \quad 
 \alpha^-=\frac{1}{p^+}\alpha^a p_a
 \;.
\end{align}
The conjugated pairs $(\bar\alpha^-,\alpha^+)$ and $(\bar\alpha^+,\alpha^-)$ are thus eliminated as independent phase-space coordinates, highlighting the fact that the only independent physical oscillators are the transverse ones $(\bar\alpha^a,\alpha^a)$. They produce states with positive norm upon quantization, as can be inferred by promoting their Poisson brackets to commutation relations
\begin{equation} \label{ab}
 [\bar\alpha^a,\alpha^b
 ]=\delta^{ab}\;,
\end{equation}
which are realized on a Fock space, where $\alpha^a$ act as creation operators while  $\bar\alpha^a$ as destruction operators, 
thus yielding a unitary spectrum of massless particles that decompose into irreps of the little group $SO(D-2)$. 

To conclude this section, we report the (partially) gauge-fixed worldline Lagrangian 
\begin{equation}
 L= p_- \dot{x}^- + p_a \dot{x}^a - \frac{1}{2p^+} p^ap^a 
 -i \bar\alpha_a \dot{\alpha}^a 
 - a \, \bar{\alpha}_a \alpha^a \;.
\end{equation}
At this stage, it only remains to address the further constraint related to the worldline gauge field~$a(\tau)$, but this has no relevance to the no-ghost theorem.

\subsection{Mass from dimensional reduction}
In this work, our main interest is to describe \emph{massive} spinning particles. One way to introduce the mass consists 
of the dimensional reduction of a higher-dimensional massless theory. We take the theory \eqref{massless bosonic action} to live in $(D+1)$-dimensions and gauge the direction $x^D$ by imposing the first-class constraint
\begin{equation}
 p_D=m\;,
\end{equation}
with $m$ the mass of the particle. We further define $(\beta, \bbeta) \coloneqq (\alpha^D, \balpha^D)$, which inherit the following Poisson brackets
\begin{equation}
\{\beta, \bar \beta\}_{\mathrm{PB}} = i \;.
\end{equation}
The constraints \eqref{constraints} get modified by the presence of the mass:\footnote{From now on, we take spacetime indices to run as $\mu=0,\dots,D-1$ where $D$ denotes the number of spacetime dimensions.}
\begin{gather}  \label{new constraints}
 H = \frac{1}{2}(p^\mu p_\mu + m^2) \;, \quad L = \alpha ^\mu p_\mu +\beta m\;, \quad \bar{L} = \bar{\alpha}^\mu p_\mu 
 +\bar \beta m\;, \quad J_c=\alpha^\mu \bar{\alpha}_\mu + \beta \bbeta - c\;,
\end{gather}
where we redefined the $U(1)$ constraint as $J_c = J - c $ for future convenience. The constant $c$
is sometimes called Chern-Simons (CS) coupling. Note that, importantly, they still satisfy the first-class algebra \eqref{algebra} despite the mass improvement. The gauge transformations \eqref{gauge trans0} are enriched by
\begin{subequations} \label{gauge trans}
\begin{align}
\delta \beta &= i \xi m + i \phi \beta\;, \label{gauge trans1} \\
\delta \bar{\beta} &= -i \bar{\xi} m - i \phi \bar{\beta}\;.
\end{align}
\end{subequations}
\paragraph{Lightcone analysis}
The lightcone gauge is implemented just as in the massless case: 
in particular, the bosonic supersymmetries are used to fix $\alpha^+= \bar\alpha^+=0$ once again, and the constraints $L=\bar L=0$ are solved by 
\begin{align}
 \alpha^-=\frac{1}{p^+}\left(\alpha^a p_a+m\beta\right)\;, \quad \bar{\alpha}^-=\frac{1}{p^+}\left(\bar\alpha^a p_a+m\bar\beta\right)\;,
\end{align}
thus eliminating the longitudinal oscillators. 
Differently from the massless case, the presence of the extra pair of $\beta$-oscillators produces a sum of irreps\footnote{This can be explicitly seen by implementing the $J_c$ constraint {\it à la} Dirac \cite{Bastianelli:2014lia}.} of the $SO(D-2)$ group that fill irreps of the $SO(D-1)$ rotation group, corresponding to the polarizations of massive spin particles in $D$ spacetime dimensions.

\subsection{Dirac quantization}
Upon covariant quantization, the worldline coordinates obey the following commutation relations fixed by their classical Poisson brackets
\begin{gather}\label{canon comm}
 [x^\mu, p_\nu]= i \delta^{\mu}_{\nu}\;, \quad [\bar{\alpha}^\mu, \alpha^\nu]= \eta ^{\mu \nu}\;, \quad [\bar{\beta}, \beta]= 1\;,
\end{gather}
and the first-class algebra becomes
\begin{equation}\label{op algebra}
 [\bar{L}, L] = 2H \;, \quad [J_c, L] = L\;, \quad [J_c, \bar{L}] = -\bar{L}\;. 
\end{equation}
Note that ordering ambiguities emerge only for the constraint $J_c$.
We have defined the quantum $J_c$ operator by a symmetric quantization prescription, so that
\begin{align}
J_c  &= \frac12 (\alpha_\mu \bar{\alpha}^\mu + \bar{\alpha}^\mu  \alpha_\mu + \beta \bbeta + \bbeta \beta) -c\;,
\cr
&=  \alpha_\mu \bar{\alpha}^\mu + \beta \bbeta  +  \frac{D+1}{2} -c\;,
\cr
&= N_{\alpha} + N_{\beta} - s\;,
\label{quantum Jc}
\end{align}
where we have used the commutation relations and introduced the usual number operators 
\be
N_{\alpha} =  \alpha_\mu \bar{\alpha}^\mu \;,
\quad  N_{\beta}=\beta \bbeta\;,
\ee
and related the CS coupling $c$ to the real number $s$ by setting
\be 
c= \frac{D+1}{2} + s  \;.
\label{2.25}
\ee
The algebra \eqref{op algebra} is now easily obtained. 
At this point, it is worth mentioning that the relation between the CS coupling $c$ and the physical value of the spin $s$ generally depends on the quantization scheme adopted.

The Hilbert space $\mathcal{H}_{\mathrm{matter}}$ of the ``matter" sector, i.e. the one  associated with the $(x,p)$ coordinates and $(\alpha,\bar\alpha,\beta,\bar\beta)$ oscillators, is realized as a tensor product of the representations of the algebras \eqref{canon comm}
\begin{equation} \label{Hm}
\mathcal{H}_{\mathrm{matter}}=\mathcal{H}_{\mathcal{M}}\otimes \mathcal{H}_{(\alpha,\beta)}\;.
\end{equation}
Specifically, we represent it by identifying the states in $\mathcal{H}_{\mathcal{M}}$ as the smooth functions of $x^\mu$, while we construct $\mathcal{H}_{(\alpha,\beta)}$ as the Fock space with vacuum defined by 
\begin{equation}
 (\bar\alpha^\mu,\bar\beta)\,\ket{0}=0\;.
\end{equation}
The decomposition of a generic state $\ket{\varphi}$ is thus written in terms of coefficients corresponding to rank-$s$ symmetric tensors:
\begin{equation}
 \ket{\varphi} = \sum_{r,p = 0}^{\infty} \ket{\varphi^{(r,p)}}= \sum_{r,p = 0}^{\infty} \frac{1}{r! p!} \, \varphi^{(r,p)}_{\mu_1 ... \mu_r}(x) \otimes \alpha ^{\mu_1} \dots \alpha ^{\mu_r} \beta^p \ket{0}\;. 
\end{equation}
The quantization may proceed either following a procedure {\it à la} Dirac or by using BRST techniques; either way, we can deal with the gauge symmetries without abandoning manifest covariance, obtaining at last a positive-definite physical Hilbert space. We start with the former method, leaving the BRST analysis for the dedicated section.\\

The physical Hilbert space in the Dirac (also known as Dirac-Gupta-Bleuler) scheme is determined by asking the constraints to have null matrix elements for arbitrary physical states  
$|\varphi\rangle$ and $|\chi\rangle$  
\begin{equation}
 \braket{\chi |(H, L, \bL, J_c)|\varphi} = 0\;.
\end{equation}
This can be satisfied by requiring 
\begin{equation}\label{phys cond Dirac}
 H\ket{\varphi} = \bL\ket{\varphi} = J_c \ket{\varphi} = 0
\end{equation}
for any physical state  $|\varphi\rangle$,
since then also $\langle \varphi | \bar L = 0 $, 
as $\bL$ is the hermitian conjugate of $L$. 

Recalling now that at the quantum level 
\be
J_c= N_{\alpha} + N_{\beta} - s
\ee
where the number operators $N_{\alpha}$ and $N_{\beta}$
count the occupation number of the $\alpha$ and $\beta$ oscillators
in the Fock space, we see 
that the quantum $J_c$ constraint selects precisely states with occupation number $s$, which must be 
a nonnegative integer.
Incidentally, we notice that the CS coupling must be quantized to get a nontrivial solution of the constraint, and therefore a nontrivial quantum theory. 
The upshot is that the quantum $J_c$ constraint
reduces the Hilbert space $\mathcal{H}_{\mathrm{matter}}$ to the subspace with occupation number $s$ for the oscillators. 

The condition \eqref{phys cond Dirac} defines physical states which are seen to form equivalence classes
\begin{equation}
 \ket{\varphi} \sim \ket{\varphi} + \ket{\varphi_{\rm null}}\;,
\end{equation}
where  $\ket{\varphi_{\rm null}}$ is  a null state of the form 
\begin{gather}
\ket{\varphi_{\rm null}} = L \ket{\xi}  \;, \qquad \text{with} \quad 
H \ket{\xi} = \bL \ket{\xi} =  (J_c+1) \ket{\xi} = 0 \;.
\end{gather}
These null states are physical, but have zero norm and vanishing overlap with any other physical state.
They give rise to redundancies or residual ``gauge symmetries" of the state $\ket{\varphi}$. 

Let us make the conditions for the case $s=1$, which is of interest to us, explicit. 
A generic state at occupation number $s=1$ is given by
\begin{equation}
 \ket{\psi} = W_\mu(x) \alpha ^\mu \ket{0} -i \varphi(x) \beta \ket{0}
\end{equation}
and the physicality conditions \eqref{phys cond Dirac} translate into the following set of equations, denoting $\Box := \partial^\mu \partial_\mu$,
\begin{align}
 (\Box - m^2) W_\mu &= 0\;,  \label{box proca dirac}\\ 
 (\Box - m^2) \varphi &= 0\;,  \\
 \partial^\mu W_\mu + m \varphi &= 0 \;, \label{trans proca Dirac}
\end{align}
with gauge symmetries related to  null states  given by
\begin{gather}
 \delta W_\mu = \partial_\mu \xi\;, \quad \delta \varphi = - m\xi\;.
\end{gather}
Using the gauge symmetry to set $\varphi(x)=0$, we recover the standard  Fierz-Pauli equations for a massive spin-1 field $W_\mu(x)$. 

\subsection{Counting degrees of freedom \label{degrees of freedom}}
In this section, we aim to use the path integral to count the number of degrees of freedom propagated by the massive model for different values of the CS coupling. 
As a byproduct, this will provide the overall normalization of the effective action we intend to study in later sections.

To count the number of degrees of freedom, we consider the one-loop effective action obtained by 
path integrating the free action on worldlines with the topology of a circle $S^1$ (the loop). After fixing 
the overall normalization to match the scalar case, we will get the number of degrees of freedom in the other
sectors of the worldline theory.

Thus, we consider the following path integral
\begin{align}
\Gamma= \int_{S^1} \frac{DG DX}{\rm Vol (Gauge)}\, \mathrm{e}^{iS[X,G]} \;,
\label{2.39}
\end{align}
where $G=(e,\bar u, u, a)$ denotes the gauge fields, whereas $X= (x^\mu ,p_\mu, \alpha^\mu, \bar \alpha_\mu, \beta, \bar \beta)$ collectively denotes all dynamical variables parametrizing the phase space. 
The action is similar to the one in \eqref{massless bosonic action} with 
the additional $(\beta,\bar\beta) $ oscillator and with
the constraints in \eqref{new constraints}, namely
\be
\begin{aligned}
S=\int  \diff\tau \biggl [ & p_\mu \dot{x}^\mu -i \bar\alpha_\mu \dot{\alpha}^\mu -i \bar\beta \dot{\beta}
-\frac{e}{2}(p^\mu p_\mu + m^2) 
- \bar{u} ( \alpha ^\mu p_\mu +\beta m)
 - u  (\bar{\alpha}^\mu p_\mu  +\bar \beta m)
 \cr &
 - a (\alpha^\mu \bar{\alpha}_\mu + \beta \bbeta - c) \biggr ] \;.
 \end{aligned}
\ee
Periodic boundary conditions are understood to implement the path integral on the circle.

The overcounting from summing over gauge equivalent configurations is formally taken into account by dividing by the volume of the gauge group. We use the Faddeev-Popov method to extract the latter, and gauge-fix the worldline gauge fields to constant moduli 
\begin{equation}\label{gauge fixed G}
 G=(e,\bar u, u, a) \ \ \to\ \ \hat{G}=(2T, 0, 0, \theta)\;.
\end{equation}
Here $T$ is the so-called ``Schwinger proper time", the modulus related to the einbein $e(\tau)$, corresponding to the gauge-invariant worldline length $\int_0^1 \diff\tau \, e$. 
The modulus $\theta\in[0,2\pi]$ is associated with the worldline $U(1)$ gauge field $a(\tau)$ and parametrizes the gauge invariant Wilson loop $\mathrm{e}^{-i \int_0^1\diff\tau \, a}$. 
It is responsible for the reduction of the Hilbert space to a given spin sector. On the other hand, the gauge fields $(u,\bar u)$ can be gauge-fixed to zero:
 this value can always be reached by inverting the differential operator that relates these fields to their respective gauge parameters  $(\xi,\bar \xi)$, as shown in Eqs. \eqref{u} and \eqref{bar u}.
 This inversion fails only at the point $\theta =0$, which, however, can be handled through a limiting procedure, as in the standard 
  $\mathcal{N}=2$ particle case \cite{Bastianelli:2005vk, Bastianelli:2005uy}.  Therefore, $(u,\bar u)$  carry no moduli.
As a reminder, moduli generically parametrize gauge-invariant field configurations that must be integrated over in the path integral.
  
We prefer to work in the Euclidean version of the theory, so that we first pass to configuration space by eliminating the momenta $p_\mu$,
Wick rotate the action with $\tau \to -i\tau$, taking into account also the rotation of the gauge field $a\to ia$, and we get the Euclidean worldline action
\begin{equation} \label{Ecugfaction}
S_{\mathrm{E}}[X,\hat{G}]=\int  \diff\tau \left [ \frac{1}{4T} \dot x^2 + \alpha_\mu (\partial_\tau + i\theta) \balpha^\mu 
+ \beta (\partial_\tau +i\theta) \bbeta 
+m^2T -i c \theta  \right ]\;.
\end{equation}
The final expression of the worldloop path integral can thus be recast as
\begin{align} \label{2.43}
\Gamma =  - \int_0^{\infty} \frac{\diff T}{T} \mathrm{e}^{-m^2 T}
\int
\frac{\diff^D\bar{x}}{(4\pi T)^{\nicefrac{D}{2}}}\ {\rm DoF}(c,D)\;,
\end{align}
where we extracted the dependence on the zero modes $\bar{x}^\mu$ of the coordinates by setting 
\begin{equation}
    x^\mu(\tau)=\bar{x}^\mu+t^\mu(\tau)\;, \quad \text{with} \quad t^\mu(0)=t^\mu(1)=0\;,
    \label{split}
\end{equation}
with the quantum fluctuations $t^\mu(\tau)$ satisfying Dirichlet boundary conditions (DBC), evaluated the free path integral that produces functional determinants, and denoted by 
DoF($c,D$)  the number of (complex) degrees of freedom, that acquires the expression
\begin{equation} \label{2.45}
{\rm DoF}(c,D) = k \int_0^{2\pi} \frac{\diff\theta}{2\pi }\, \mathrm{e}^{i c \theta}\;
{\rm Det}\left(\partial_\tau -i\theta\right) 
{\rm Det}\left(\partial_\tau +i\theta\right) 
\left[{\rm Det}(\partial_\tau +i\theta)\right]^{-D-1}\;,
\end{equation}
with $k$ an overall normalization to be fixed later on. 
The value DoF $=1$ corresponds to a complex scalar, as seen by comparing with QFT expressions.
In this formula, the first two functional determinants are the Faddeev-Popov ones,
whereas the third one is due to the path integration over the bosonic oscillators. All determinants are evaluated with periodic boundary conditions:
using 
\begin{equation}\label{det}
{\rm Det}\left(\partial_\tau + i\theta\right) = 2i \sin \left(\frac{\theta}{2}\right)\;,
\end{equation}
see for example \cite{Bastianelli:2007pv, Dai:2008bh}, 
setting the CS coupling to\footnote{The shift from the value given in \eqref{2.25} is due to the contribution of the ghost fields. For convenience, we now indicate the degrees of freedom by 
${\rm DoF}(s,D)$, which highlights the dependence on the value of the spin $s$, rather than on the CS coupling $c$.
This should not cause any confusion.}
$c = \frac{D - 1}{2} + s$, and fixing $k=- 1$ as overall normalization, we find the following expression for the number of degrees of freedom
\begin{align}
{\rm DoF}(s,D) &=  \int_0^{2\pi} \frac{\diff\theta}{2\pi }\; \mathrm{e}^{i \left(\frac{D-1}{2}+s\right) \theta} \;\left(2i \sin \frac{\theta}{2}\right )^{1-D}\;.
\end{align}
To evaluate it, we find it more convenient to recast it in terms of the Wilson loop variable
$w:=\mathrm{e}^{-i\theta}$, so that
\begin{equation}
{\rm DoF}(s,D) 
= 
\oint \frac{\diff w}{2\pi i}\, \frac{1}{w^{s + 1}}\, \frac{1}{(1-w)^{D-1}}\;.
\end{equation}
Deforming the contour to exclude the singular point $w=1$, while taking care of the pole in $w=0$, we get
\begin{align}
\begin{split}
&{\rm DoF}(0,D) = 1\;, \\ 
&{\rm DoF}(1,D) = ( D-1)\;, \\
&{\rm DoF}(2,D) =  \frac{D(D-1) }{2}\;, \\
& \cdots \\
&{\rm DoF}(s,D) =\frac {(D-1) D\cdots (D+s-2)}{s!}\;,
\end{split}
\end{align}
which indeed describes the degrees of freedom of a reducible (for $s \geq 1$) representation of the little group $SO(D-1)$
as carried by a symmetric tensor with $s$ indices. It corresponds  to the propagation of a 
  multiplet of massive particles of decreasing 
 spin $s, s-2, s-4, \cdots, 0$ for even $s$, and $s, s-2,\cdots, 1$ for odd $s$. 
 This matches the results seen in the lightcone gauge.

\section{Coupling to electromagnetism} \label{sec2} 
The BRST formalism is especially well-suited for analyzing the constraints required for consistent background interactions. For this reason, we briefly review the free particle in this framework and then examine its interaction with an electromagnetic background.

\subsection{Free BRST analysis} \label{sec2.1}
We proceed with the BRST quantization focusing only on the subalgebra of \eqref{op algebra} generated by $(H, L, \bar{L})$. The constraint $J_c$ is treated on different footings: it is imposed as a constraint on the
BRST Hilbert space, defining a restricted Hilbert space where the cohomology of the BRST operator will be analyzed.

The Hilbert space is enlarged to realize the fermionic ghost-antighost pairs of operators 
\begin{gather}
 \{b, c\} = 1 \;, \quad \{\mathcal{B},\bar{\mathcal{C}}\} = 1 \;, \quad \{\bar{\mathcal{B}},\mathcal{C}\} = 1\;,
\end{gather}
associated with the $(H, L, \bar{L})$ constraints, respectively. We assign them the following ghost numbers: $\mathrm{gh}(c, \bar{\mathcal{C}},\mathcal{C} ) = +1\,, \;\mathrm{gh}(b, \bar{\mathcal{B}},\mathcal{B} ) = -1$. The BRST charge associated with a first-class system is readily constructed. In the present case, it takes the form
\begin{equation} \label{Q}
 \mathcal{Q} = cH + \bar{\mathcal{C}} L + \mathcal{C} \bar{L} -2\mathcal{C} \bar{\mathcal{C}} b \;.
\end{equation}
It is an anticommuting, ghost number $+1$, nilpotent operator by construction. It is hermitian provided that
\begin{gather}
 c^\dagger\ = c \;, \quad b^\dagger = b \;, \quad \C^\dagger = \bC \;, \quad \B^\dagger = \bB \;. 
\end{gather}
The matter sector Hilbert space \eqref{Hm} is extended to the BRST Hilbert space $\mathcal{H}_{\mathrm{BRST}}$ by a tensor product with the ghost sector, associated with the $(c,b,\mathcal{B},\bar{\mathcal{C}}, \mathcal{C},\bar{\mathcal{B}})$ operators. The latter is constructed 
as a Fock space on the ghost vacuum defined by 
\begin{equation}
(b,\bar{\mathcal{C}},\bar{\mathcal{B}})\,\ket{0}_{\mathrm{gh}}=0\;.
 \end{equation}
Since all ghosts are Grassmann odd, $\mathcal{H}_{\mathrm{gh}}$ is finite dimensional.\footnote{In particular, the Fock vacuum $\ket{0}_{\mathrm{gh}}$ can be mapped into the ``physical vacuum" $\ket{1}_{\mathrm{gh}}$, with the concept of ``physicality" to be defined shortly,  
by
\begin{equation*}
 \ket{1}_{\mathrm{gh}}:=\mathcal{B}\ket{0}_{\mathrm{gh}}\ ,
\end{equation*} 
see e.g. the discussion in \cite{Bengtsson:2004cd, Bonezzi:2024emt}. The vacuum $\ket{1}_{\mathrm{gh}}$ turns out to be the correct one to consider in order to create external states by inserting vertex operators in the worldline path integral, as illustrated in \cite{Bastianelli:2025xx}.} A generic state $\ket{\Phi}$ in the BRST-extended Hilbert $\mathcal{H}_{\mathrm{BRST}}$ space reads
\begin{gather}\label{gen state}
 \ket{\Phi} = \sum_{s,p = 0}^{\infty} \sum_{q,r,t = 0}^{1} c^q \mathcal{C}^r \mathcal{B}^t\ket{\Phi^{(s,p)(q,r,t)}} 
\end{gather}
where
\begin{gather}
 \ket{\Phi^{(s,p)(q,r,t)}} = \frac{1}{s! p!} \, \Phi^{(s,p)(q,r,t)}_{\mu_1 ... \mu_s}(x)\, \alpha ^{\mu_1} \dots \alpha ^{\mu_s} \beta^p \ket{0}\;,
\end{gather}
with $\ket{0}$ now denoting the full BRST vacuum. With this choice, the conjugate momenta act as derivatives:
\begin{equation}
 p_\mu=-i\partial_\mu\;, \quad \bar\alpha^\mu=\partial_{\alpha^\mu}\;, \quad \bar\beta=\partial_{\beta}\;, \quad b=\partial_c\;, \quad \bar{\mathcal{C}}=\partial_\mathcal{B}\;, \quad \bar{\mathcal{B}}=\partial_{\mathcal{C}}\;.
\end{equation}
We now introduce a couple of operators, $G$ and $\J_s$, to further restrict the full BRST Hilbert space. These are the ghost number operator
\begin{align} 
 G &= cb + \mathcal{C} \bar{\mathcal{B}} - \mathcal{B} \bar{\mathcal{C}}\;, \qquad [G,\Q] = \Q \;,
\end{align}
and the (shifted) occupation number operator $\J_s$
\begin{align}
 \J_s &= \alpha_\mu \bar{\alpha}^\mu + \beta \bar{\beta} + \mathcal{C} \bar{\mathcal{B}} + \mathcal{B} \bar{\mathcal{C}} - s\;, \quad [\Q,\J_s] = 0\;.
\end{align}
They commute between themselves, $[G,\J_s]=0$.  
The ghost number operator grades the BRST Hilbert space according to the ghost number, and the commutator $ [G,\Q] = \Q$ manifests that the BRST charge has ghost number 1.
The occupation number operator  $\J_s$ also grades the Hilbert space according to its eigenvalues and
can be used as a constraint to project the Hilbert space onto the subspace with fixed occupation number $s$.\footnote{We choose an antisymmetric quantization's prescriptions for fermionic operators. In combination with the Weyl ordering for the bosonic ones previously discussed, the effect is to shift the CS coupling as $c = \frac{D - 1}{2} + s$.
This relation has already been used in the path integral construction (see footnote 4), which evidently involves a regularization consistent with this ordering prescription.} 

We can exploit the operators above simultaneously -- since $[G, \J_s] = 0$ -- to select states in $\mathcal{H}_{\mathrm{BRST}}$ with a precise ghost and occupation number. The physical states are identified as elements of the BRST cohomology
\begin{gather} \label{cohomology}
 \Q \ket{\Phi} = 0\;,  \quad \ket{\Phi} \sim \ket{\Phi} + \Q \ket{\Lambda} 
\end{gather}
restricted to the subspace with vanishing eigenvalues of the ghost number and shifted occupation number operators, i.e.
\begin{gather}
 G\ket{\Phi}=\J_s \ket{\Phi} = 0\;.
\end{gather}

Our interest lies in the first-quantized description of a massive spin-1 particle: this is achieved by choosing $s=1$. 
An arbitrary wavefunction  at zero ghost number and with $s=1$ is then given by
\begin{equation}\label{eq: s=1 state}
 \ket{\psi} = W_\mu(x) \alpha ^\mu \ket{0} -i\varphi (x) \beta \ket{0} +f(x) c \mathcal{B} \ket{0}\;,
\end{equation}
where the complex fields $W_\mu(x)$, $ \varphi (x) $, and $f(x)$ must be further constrained by Eq.~\eqref{cohomology} to represent the physical states of the theory.
From the closure equation, i.e., the first one in \eqref{cohomology}, we obtain
\begin{subequations}\label{Proca0}
\begin{align}
\left( \Box-m^2 \right)W_\mu-2i\partial_\mu f&=0\;, \\
\left( \Box-m^2 \right)\varphi+2imf&=0\;, \\
\partial_\mu W^\mu+m\varphi-2if&=0\label{algebraic}\;,
\end{align}
\end{subequations}
which, upon eliminating the auxiliary field $f(x)$, represent the field equations of the Proca field in the St\"uckelberg formulation
\begin{subequations}\label{Proca}
\begin{align}
\left( \Box-m^2 \right)W_\mu-\partial_\mu\partial \cdot W
-m\partial_\mu\varphi&=0\;, \\
\Box \varphi+m\partial_\mu W^\mu&=0\;,
\end{align}
\end{subequations}
where the dot ``$\cdot$" indicates contraction over spacetime indices. The latter equations enjoy a gauge symmetry, which, from \eqref{cohomology}, reads
\begin{equation}
 \delta\ket{ \psi} = Q \ket{\Lambda}\;, \quad \text{with} \quad \ket{\Lambda} = i \xi(x) \mathcal{B} \ket{0}\;,
\end{equation}
i.e.
\begin{equation}\label{gauge}
 \delta W_\mu=\partial_\mu\xi\;, \quad \delta\varphi=-m \xi\;,
\end{equation}
which is the well-known St\"uckelberg gauge symmetry.

A few comments are in order: (\textit{i}) the wavefunction \eqref{eq: s=1 state} can be interpreted as a spacetime Batalin-Vilkovisky (BV) ``string field" displaying only
the classical fields out of the minimal BV spectrum of the Proca theory, along with an auxiliary field.\footnote{The complete minimal BV spectrum is obtained by relaxing the condition $G\ket{\psi}=0$, see for instance \cite{Carosi:2021wbi, Fecit:2023kah}.} The Grassmann parities and ghost numbers of the field components are all equal to zero.
Note the presence of the St\"uckelberg scalar $\varphi$, which restores the $U(1)$ gauge symmetry \cite{Stueckelberg:1957zz}, originally broken due to the introduction of the mass.
(\textit{ii}) In the so-called \emph{unitary gauge}, namely setting the St\"uckelberg field to zero, one reduces the field equations to the standard Fierz-Pauli system for the massive spin-1 field $W_\mu(x)$. (\textit{iii}) Taking the massless limit produces from \eqref{Proca} a pair of decoupled equations: one for a free-propagating massless vector field $W_\mu(x)$ and one for a massless scalar field $\varphi(x)$. This is tantamount to the fact that the theory of massive spin-1 does not suffer from the so-called ``vDVZ discontinuity", differently from the massive spin 2 case \cite{vanDam:1970vg, Zakharov:1970cc}.

\subsection{Consistent electromagnetic coupling}\label{coupling}
The coupling of the worldline to an abelian background field $A_\mu(x)$ in spacetime (with coupling constant $q$) is achieved by covariantizing the $(L,\bar L)$ constraints as follows
\begin{gather}
\label{cov constr}
 L \rightarrow \alpha^\mu \pi_\mu + \beta m\;, \quad \bL \rightarrow \balpha^\mu \pi_\mu + \bbeta m\;,
\end{gather}
where the covariantized momentum $\pi_\mu$ 
with coupling constant $q$ is defined by
\begin{equation}
\pi_\mu = p_\mu -q A_\mu\;.
\end{equation}
It becomes the covariant derivative in the coordinate representation, $\pi_\mu= -i(\partial_\mu -iq A_\mu)=-i D_\mu$.
The new constraints do not form a first-class algebra anymore: while the bosonic supersymmetry charges do commute into a possibly deformed Hamiltonian
\begin{align}
 [\bL, L] &= \pi^2 + m^2 +\balpha^\mu \alpha^\nu \tilde{F}_{\mu\nu}=:2 H_{- 1/2}\;,
\end{align}
where we have denoted $[\pi_\mu, \pi_\nu]=-[D_\mu, D_\nu]=  iq F_{\mu\nu}=:\tilde{F}_{\mu\nu}$, we find that the remaining commutators read
\begin{align}
 [L, H_{-1/2}] &= i \alpha^\mu \pa^\nu \tilde{F}_{\nu \mu} + \frac{3}{2}\alpha^\mu \tilde{F}_{\mu \nu} \pi^\nu +\frac{i}{2} \balpha^\nu \alpha^\rho \alpha^\mu \pa_\mu \tilde{F}_{\rho \nu} \;, \\
 [\bL, H_{-1/2}] &=i \balpha^\mu \pa^\nu \tilde{F}_{\nu \mu} + \frac{3}{2}\balpha^\mu \tilde{F}_{\mu \nu} \pi^\nu +\frac{i}{2} \balpha^\nu \alpha^\rho \balpha^\mu \pa_\mu \tilde{F}_{\rho \nu}\;,
\end{align}
and do not allow for a suitable redefinition of the constraints to form a first-class algebra. Thus, we expect the associated BRST charge to fail to be nilpotent, which indicates an inconsistency of the interacting worldline theory at the quantum level. We then proceed tentatively and try an ansatz
\begin{equation}\label{Q defor}
 \mathcal{Q}_A = c H_\kappa + S^\mu \pi_\mu + \bar{\mathcal{C}}\beta m + \mathcal{C} m \bar{\beta} - Mb\;,
\end{equation}
with a deformed Hamiltonian
\begin{equation}\label{Hdef}
 H_\kappa =\frac{1}{2}\left(\pi ^2 + m^2 + 2 \kappa \alpha ^\mu \balpha^\nu \tilde F_{\mu\nu} \right) 
\end{equation}
that contains a non-minimal coupling with constant $\kappa$ to be conveniently fixed, and then compute
\begin{equation}\label{Q2}
 \mathcal{Q}_A^2 =-\frac{2 \kappa + 1}{4} M S^{\mu \nu} \tilde{F}_{\mu \nu} +c [H_\kappa, S^\mu \pi_\mu] \;,
\end{equation}
where we used the shorthand notations
\begin{gather}
 S^\mu = \alpha^\mu \bar{\mathcal{C}} + \bar{\alpha}^\mu \mathcal{C}\;, \quad S^{\mu \nu} = \alpha^\mu \bar{\alpha}^\nu - \alpha^\nu \bar{\alpha}^\mu\;, \quad M = 2 \mathcal{C} \bar{\mathcal{C}}\;.
\end{gather}
In general, \eqref{Q2} is not zero, except for the trivial case of vanishing field strength, which manifests the inconsistency of coupling massive spin $s$ particles, with generic $s$, to an EM background. This is also the case for massless particles, as already discussed in \cite{Bonezzi:2024emt}. However, restricting the occupation number to be $s\leq1$, the nilpotency condition simplifies:\footnote{This can be inferred by counting the number of annihilation operators in $ \mathcal{Q}_A^2$: if there are two or more, they annihilate the physical wavefunction for $s=1$.}
\begin{align}
\begin{split} \label{Q^2}
 \mathcal{Q}_A^2\big\vert_{s=0,1} &= c [H_\kappa, S^\mu \pi_\mu]\big\vert_{s=0,1}\\
 &=-\frac{i c}{2}\left(\partial_\mu \tilde{F}^{\mu \nu} S_\nu + 2 i(1 -\kappa) \tilde{F}^{\mu \nu} \pi_\mu S_\nu -\kappa \, S^\nu S^{\alpha \beta} \partial_\nu \tilde{F}_{\alpha \beta}\right)\big\vert_{s=0,1}\;.
\end{split}
\end{align}
For the $s=0$ sector, this expression is automatically zero regardless of any condition on the background electromagnetic field, as this operator contains destruction operators sitting on the right that annihilate the $s=0$ wave function (recall the expressions for the operators $S^\mu$ and $S^{\mu\nu}$).
Physically, this expresses the fact that spinless particles can be consistently coupled to off-shell abelian background fields. 
As for the massive spin-1 sector, cf. \eqref{eq: s=1 state}, the previous equation further simplifies to
\begin{equation}
 \mathcal{Q}_A^2\big\vert_{s=1}=\frac{q}{2} c(\alpha^\nu \bar{\mathcal{C}} - \bar{\alpha}^\nu \mathcal{C}) \partial^\mu (\partial_\mu A_\nu - \partial _\nu A_\mu)\;,
\end{equation}
having set $\kappa=1$ to achieve this result: then, nilpotency of the deformed BRST charge requires the background $A_\mu(x)$ to be on-shell, i.e.
\begin{equation}
 \partial^\mu F_{\mu\nu}=\partial^\mu(\partial_\mu A_\nu - \partial_\nu A_\mu) \must 0\;.
\end{equation}
This is enough to prove the consistency of the coupling.

Let us notice that the mass does not obstruct the nilpotency, namely, it does not seem to carry substantial differences with respect to the massless case. To be more precise, $m$ does not \emph{explicitly} enter in the BRST algebra for any spin $s$, but it may obstruct the nilpotency for higher-spin particles, starting from the spin 2 case as discussed in \cite{Fecit:2023kah} for the gravitational coupling. 
How the no-go theorem about massless charged particles \cite{Weinberg:1980kq} appears from a worldline perspective is at the moment unclear to us.

\section{Effective action in electromagnetic background}\label{sec3}
In this section, we employ the worldline model to compute the one-loop effective action induced by a charged spin-1 particle in a constant electromagnetic background.

The worldline representation of the effective action is derived by following the same approach as in the free case (cf. Section \ref{degrees of freedom}), which in particular determines the overall normalization of the path integral. 
Given the BRST analysis presented in Section \ref{coupling}, we are naturally led to treat the $s=0$ and $s=1$ cases simultaneously. 
This approach allows for a direct comparison, with the spinless case serving as a check on the novel spin-1 contribution within the first-quantized framework.

As the interacting worldline action, we take the covariantized version of the gauge-fixed free action in Euclidean configuration space \eqref{Ecugfaction}
with covariantized constraints \eqref{cov constr} and deformed Hamiltonian $H_{1}$ \eqref{Hdef}, i.e. (factoring out the $m^2T -i c \theta$ constant term)
\begin{equation}
\begin{split}
 S_{\mathrm{E}}[X, \hat{G}; A] = \int \diff \tau \left[\frac{\dot{x}^2}{4T} - i q A^\mu \dot{x}_\mu + \alpha^\mu \left( \delta_{\mu \nu} \left(\frac{\diff}{\diff\tau}+ i \theta \right) + 2 i q T F_{\mu \nu} \right) \balpha^\nu + \beta \left(\frac{\diff}{\diff\tau} + i \theta \right) \bbeta 
\right]\;.
\end{split}
\end{equation}
We restrict our analysis to four spacetime dimensions and consider a constant electromagnetic field as the on-shell background. Under these conditions, we derive the one-loop effective action of the Euler–Heisenberg type induced by a massive spin-1 particle. This effective action is given 
by the path integral on the circle of the gauge-fixed action and takes the form
\begin{equation}\label{path integral}
 \Gamma[A] =  \int_{0}^{\infty} \frac{\diff T}{T} \mathrm{e}^{-m^2 T} 
 \int_{0}^{2\pi} \frac{\diff\theta}{2\pi} 
 \mathrm{e}^{i c \theta} 
 \,{\rm Det}\left(\partial_\tau -i\theta\right) {\rm Det}\left(\partial_\tau +i\theta\right) \int_{\mathrm{PBC}} DX \,\mathrm{e}^{-S_\mathrm{E}[X,\hat{G}; A]} \;,
\end{equation}
with measure in moduli space and determinants already fixed by the free case, see Eqs. \eqref{2.43} and \eqref{2.45}, and
with the CS coupling fixed to  $c=\frac32 +s$.
Recalling the coordinate split in Eq.~\eqref{split},
 we use the Fock-Schwinger gauge around $\bar{x}$ for the background field, i.e.
\begin{equation}\label{FS}
 (x-\bar{x})^\mu A_\mu (x)=0\;,
\end{equation}
to express derivatives of the gauge potential at the point $\bar{x}$ in terms of derivatives of the field strength tensor
\begin{equation}\label{FS2}
 A_\mu(\bar{x}+t)=\frac12 t^\nu F_{\nu\mu}(\bar{x})+\dots\;,
\end{equation}
where the higher-derivative terms hidden inside the dots vanish since we focus on the constant electromagnetic background case. 
Then, the path integral becomes Gaussian, and it simplifies to
\begin{equation}
\begin{split}\label{path integral 2}
 \Gamma[A]& = \int\diff^4 \bar{x} \int_{0}^{\infty} \frac{\diff T}{T} \frac{\mathrm{e}^{-m^2 T}}{(4\pi T)^2} \int_{0}^{2\pi} \frac{\diff \theta}{2\pi} \, \mathrm{e}^{i\left(\frac{3}{2}+ s\right) \theta} \;4 \sin^2\left(\frac{\theta}{2}\right)\int_{\mathrm{DBC}} Dt \, \mathrm{e}^{-S_{t}[X,\hat{G};A]} \\ 
 &\phantom{=\;} \int_{\mathrm{PBC}} D\alpha D\balpha \, \mathrm{e}^{-S_{\alpha}[X,\hat{G};A]} \int_{\mathrm{PBC}} D\beta D\bbeta \, \mathrm{e}^{-S_\beta[X,\hat{G}]}\;,
\end{split}
\end{equation} 
where we factored out the normalization of the free particle path integral, and where we defined
\begin{align}
 S_t[X,\hat{G};A] &= \int \diff\tau \, \frac{1}{2} t^\mu \Delta_{\mu\nu}^{(t)} t^\nu\;, \quad \text{with} \quad \Delta_{\mu\nu}^{(t)}=- \frac{1}{2T} \delta_{\mu \nu}\frac{\diff^2}{\diff\tau^2} - i q F_{\mu \nu}\frac{\diff}{\diff\tau}\;,\\
 S_\alpha[X,\hat{G};A] &= \int \diff\tau \, \alpha^\mu \Delta_{\mu\nu}^{(\alpha)}\balpha^\nu\;, \quad \hspace{-0.05cm}\text{with} \quad \Delta_{\mu\nu}^{(\alpha)}=\delta_{\mu\nu}\left(\frac{\diff}{\diff\tau} + i \theta \right) + 2i q T F_{\mu \nu}\;,\\
 S_\beta[X,\hat{G}] &= \int \diff\tau \beta \Delta^{(\beta)}\bbeta\;, \quad \hspace{0.45cm}\text{with} \quad \Delta^{(\beta)} =\frac{\diff}{\diff\tau} + i \theta\;,
\end{align}
in order to highlight the three differential operators whose functional determinants have to be computed as a result of the path integration over the variables $X(\tau)$. We apply the Gel’fand–Yaglom (GY) theorem \cite{Gelfand:1959nq} to compute the first one, leaving details and conventions in Appendix \ref{appA}, while the remaining two can be directly inferred from the previous result \eqref{det}. Using the required  boundary conditions, 
as  indicated in Eq.~\eqref{path integral 2},  we get
\begin{align}
 \mathrm{Det}\left(\Delta_{\mu\nu}^{(t)} \right)&= {\rm det} \left(\frac{\sin(q T F_{\mu \nu})}{q T F_{\mu \nu}}\right) \;, \label{Det1}\\
 \mathrm{Det}\left( \Delta_{\mu\nu}^{(\alpha)} \right)&= {\rm det}\left[2i \sin\left(\frac{\theta}{2}\delta_{\mu \nu} + q T F_{\mu \nu} \right)\right]\;, \label{Det3} \\
 \mathrm{Det}\left(\Delta_{\mu\nu}^{(\beta)} \right)&=2i \sin\left(\frac{\theta}{2}\right)\label{Det2}\;,
\end{align}
having already extracted the zero modes for the $x$-coordinates. Our final expression is 
\begin{equation}
 \boxed{\Gamma[A] = \int \diff^4 \bar{x} \int_{0}^{\infty} \frac{\diff T}{T}\frac{ \mathrm{e}^{-m^2 T}}{(4 \pi T)^2} \,{\rm det}^{-\nicefrac{1}{2}} \left(\frac{\sin(q T F_{\mu \nu})}{q T F_{\mu \nu}}\right)\, I_s(T, A)}\;,
\end{equation}
where all that is left to do is to perform the modular integration in $\theta$ for a given value of spin $s$:
\begin{equation}\label{I_s}
I_s(T, A) = \int_{0}^{2\pi} \frac{\diff\theta}{ 2\pi i} \, \mathrm{e}^{i\left(\frac{3}{2}+ s\right)\theta} \, 2\sin\left(\frac{\theta}{2}\right) \, {\rm det}^{-1} \left[2i \sin\left(\frac{\theta}{2}\delta_{\mu \nu} + q T F_{\mu \nu} \right)\right] \;.
\end{equation}
It is convenient to recast the determinants above by diagonalizing the (Euclidean) field strength tensor,\footnote{Explicitly: $F_{4 i} = -i E_i, \ F_{ij} = \epsilon_{ijk} B_k, \ (i,j = 1, 2, 3)$.} given that its eigenvalues are
\begin{gather}\label{diag1}
 \lambda_1 = K_- \;, \quad \lambda_2 = i K_+ \;, \quad \lambda_3 = - K_- \;, \quad \lambda_4 = -i K_+ \;,
\end{gather}
having defined $K_{\pm} = \sqrt{\sqrt{\mathcal{F} ^2 + \mathcal{G}^2} \pm \mathcal{F}}$ in terms of the Maxwell invariants
\begin{gather}\label{diag2}
 \mathcal{F} = \frac{1}{4} F_{\mu \nu} F^{\mu \nu} = \frac{\vec{B}^2 - \vec{E}^2 }{2}\;,  
  \quad \mathcal{G} = -\frac{i}{4} \tilde{F}_{\mu \nu} F^{\mu \nu} = \vec{E} \cdot \vec{B} \;.
\end{gather}
The modular integration in the Wilson variable $w=\mathrm{e}^{-i\phi}$ is then
\begin{equation}
 I_s(T, A) = \oint \frac{\diff w}{2 \pi i} 
 \frac{1}{w^{s +1} }
 \frac{w-1 }{\left(1 + w^2-2 w \mathcal{K}_+\right) \left(1 + w^2-2 w \mathcal{K}_-\right)}\;,
\end{equation} 
where $\mathcal{K_{+}}= \cosh( 2 q T K_{+})$ and $\mathcal{K_{-}}= \cos( 2 q T K_{-})$.\\

We now have all the ingredients to investigate the effective action $\Gamma[A] = \int \diff^4 \bar{x}\, \mathcal{L}[A]$
for spin $s=0,1$.
 In particular:
\begin{enumerate}[label=(\roman*)]
 \item The scalar case $s=0$, which corresponds to scalar QED, comes from the simple pole at $w=0$
 \begin{align}
I_0(T, A) &= \mathrm{Res} \left[ \frac{w - 1}{w \left(1 + w^2-2 w \mathcal{K}_+\right) \left(1 + w^2-2 w \mathcal{K}_-\right)}\right]_{w=0} =  -1\;,
 \end{align}
 hence it correctly reproduces the celebrated Weisskopf Lagrangian \cite{Weisskopf:1936hya}
 \begin{equation}
 \boxed{\mathcal{L}_{s=0}[A] =- \int_{0}^{\infty} \frac{\diff T}{T}\frac{ \mathrm{e}^{-m^2 T}}{(4 \pi T)^2} \, \frac{q^2 T^2 K_- K_+ }{\sinh(qTK_+)\sin(qTK_-) }}\;.
 \end{equation}
 
 \item The massive spin $1$ case instead arises form the double pole at $w=0$
 \begin{align}
 I_1(T, A) &= \mathrm{Res} \left[ \frac{w - 1}{w^{2} \left(1 + w^2-2 w \mathcal{K}_+\right) \left(1 + w^2-2 w \mathcal{K}_-\right)}\right]_{w=0} = 1 - 2(\mathcal{K}_+ + \mathcal{K}_-)\;,
 \end{align}
 leading to
 \begin{equation}\label{effect action}
 \boxed{\mathcal{L}_{s=1}[A] =\int_{0}^{\infty} \frac{\diff T}{T}\frac{ \mathrm{e}^{-m^2 T}}{(4 \pi T)^2} \, \frac{q^2 T^2 K_- K_+ }{\sinh(qTK_+)\sin(qTK_-) } \, \left[1- 2\cosh( 2 q T K_+) -2 \cos( 2 q T K_-) \right]}\;.
 \end{equation}
  \end{enumerate}
This last expression corresponds to the Heisenberg-Euler effective Lagrangian for a massive charged vector boson in a constant electromagnetic background. It was originally derived in 1965 by Vanyashin and Terent’ev, starting from a quantum field theory of vector electrodynamics \cite{Vanyashin:1965ple}. 
 In contrast, our derivation employs a self-consistent first-quantized approach, which offers a more direct and transparent computation than the conventional second-quantized formalism. This constitutes the main result we set out to obtain using the worldline method.
  
 Our approach offers a natural framework for exploring possible extensions. For instance, one could interpret our final expression as the result of a locally constant field approximation \cite{Fedotov:2022ely} and investigate corrections by systematically including higher-order terms in \eqref{FS2}. This would likely involve following a procedure similar to that of \cite{Fecit:2025kqb} for performing perturbative corrections from the worldline, ultimately leading to the determination of the generalized heat kernel coefficients computed in \cite{Franchino-Vinas:2023wea, Franchino-Vinas:2024wof}. We leave this analysis to future work.
 
 As a final note, let us report the perturbative expression given by an expansion in the particle's electric charge $q$ 
 \begin{equation}
 \mathcal{L}_{s=1}[A] =\int_{0}^{\infty} \frac{\diff T}{T}\frac{ \mathrm{e}^{-m^2 T}}{(4 \pi T)^2} \, 
 \left(-3 + \frac{7}{4} q^2 T^2 \, \mathrm{tr}[F_{\mu \nu}^2] + \frac{5}{32} q^4 T^4 \,\mathrm{tr}^2[F_{\mu \nu}^2] - \frac{27}{40} q^4 T^4 \, \mathrm{tr}[F_{\mu \nu}^4] + \smallO(q^6)\right)\;.
 \end{equation}
 The first two terms give divergent contributions, the first one being an infinite vacuum energy, while the second one corresponds to the one-loop divergence in the photon self-energy, and they should be renormalized away. On the other hand, the last two terms are finite and give rise to the quartic interaction’s contributions once integrated in the proper time. Thus, the leading terms of the renormalized effective (Euclidean) Lagrangian, with the tree-level Maxwell term included,
  are expressed as 
\begin{equation}
 \mathcal{L}^{\mathrm{ren}}_{s=1}[A] = \frac{1}{4} F_{\mu \nu} F^{\mu \nu} + \frac{q^4}{16 \pi^2 m^4} \left(\frac{5}{32} (F_{\mu \nu} F^{\nu \mu})^2 -\frac{27}{40} F^{\mu \nu} F_{\nu \rho} F^{\rho \sigma} F_{\sigma \mu}\right) +\cdots
\end{equation}
which shows the leading vertices for the scattering of light by light.

An overall minus sign arises upon continuation back to Minkowski spacetime.
Inserting this sign, the Lagrangian in Minkowski spacetime can be written in the more explicit form
\begin{equation}\begin{aligned}
 \mathcal{L}^{\mathrm{ren}}_{s=1}[A] &= - \frac{1}{4} F_{\mu \nu} F^{\mu \nu} + \frac{q^4}{16 \pi^2 m^4} \left(-\frac{5}{32} (F_{\mu \nu} F^{\nu \mu})^2 +\frac{27}{40} F^{\mu \nu} F_{\nu \rho} F^{\rho \sigma} F_{\sigma \mu}\right)+ \cdots
 \cr 
 &= \frac12 (\vec{E}^2 - \vec{B}^2)
 + \frac{\alpha^2}{40 m^4} \left ( 29 (\vec{E}^2 - \vec{B}^2)^2 + 108 (\vec{E}\cdot \vec{B})^2 \right ) 
 + \cdots
 \end{aligned}\end{equation}
where, for ease of comparison with the literature, we have introduced the fine-structure constant $\alpha =\frac{q^2}{4\pi}$ in natural units, and used the relations
\begin{equation}
F_{\mu \nu} F^{\mu \nu}= 2 (\vec{B}^2 - \vec{E}^2)\;, \quad F^{\mu \nu} F_{\nu \rho} F^{\rho \sigma} F_{\sigma \mu}= 2 (\vec{E}^2 - \vec{B}^2)^2 + 4 (\vec{E}\cdot \vec{B})^2\;,
\end{equation}
to obtain the second line. 

It may be interesting to compare this result with the more widely known results for the
spin-0 and spin-$\frac{1}{2}$ cases, which we include here for convenience:
\begin{equation}\begin{aligned}
 \mathcal{L}^{\mathrm{ren}}_{s=0}[A] 
&= -\frac{1}{4} F_{\mu \nu} F^{\mu \nu} + \frac{q^4}{16 \pi^2 m^4} \left( \frac{1}{288} (F_{\mu \nu} F^{\nu \mu})^2 +\frac{1}{360} F^{\mu \nu} F_{\nu \rho} F^{\rho \sigma} F_{\sigma \mu}\right) 
 + \cdots
\cr
&= \frac12 (\vec{E}^2 - \vec{B}^2)
+ \frac{\alpha^2}{360 m^4} \left ( 7 (\vec{E}^2 - \vec{B}^2)^2 + 4 (\vec{E}\cdot \vec{B})^2 \right ) + \cdots
\end{aligned}\end{equation}
and
\begin{equation}\begin{aligned}
  \mathcal{L}^{\mathrm{ren}}_{s=\frac12}[A] 
 &= - \frac{1}{4} F_{\mu \nu} F^{\mu \nu} + \frac{q^4}{16 \pi^2 m^4} \left(-\frac{1}{32} (F_{\mu \nu} F^{\nu \mu})^2 +\frac{7}{90} F^{\mu \nu} F_{\nu \rho} F^{\rho \sigma} F_{\sigma \mu}\right) + \cdots
\cr
&= \frac12 (\vec{E}^2 - \vec{B}^2)+ \frac{2 \alpha^2}{45 m^4} \left ( (\vec{E}^2 - \vec{B}^2)^2 + 7 (\vec{E}\cdot \vec{B})^2 \right ) + \cdots \;.
\end{aligned}\end{equation}
They arise from the Weisskopf and Euler–Heisenberg effective Lagrangians, respectively.
 
\subsection{Production of massive spin-1 particle pairs}\label{sec3.2}
It is well-known that if the effective action in the presence of a classical background field assumes a non-vanishing imaginary contribution, this has the physical interpretation of an instability of the quantum field theory vacuum. In turn, this signals the appearance of states with a non-vanishing number of particles, namely, a production of particle-antiparticle pairs takes place. This is the essence of the so-called ``Schwinger effect" \cite{Schwinger:1951nm}. Quantitatively, the Minkowskian effective action is related to the vacuum persistence probability by
\begin{align}
 \vert\langle 0_{\rm out} \vert 0_{\rm in}\rangle\vert^2 =\mathrm{e}^{- 2\operatorname{Im} \Gamma_\mathrm{M}}\;,
\end{align}
from which the pair production probability is given by $P_{\mathrm{pair}}:=1-\mathrm{e}^{- 2\operatorname{Im} \Gamma_\mathrm{M}} \approx 2 \, \mathrm{Im} \,\Gamma_{\mathrm{M}}$.\\
In this section, we compute the rate for the Schwinger pair production of massive charged spin-1 particles in a constant external electric field $\vec{E}$.
The effective action \eqref{effect action} with   $K_+ =0$ and $K_- =E$, with  $E$ being the modulus of the electric field,
reduces to
\begin{equation}\label{effect action E B}
 \Gamma[A] =\int \diff^4 \bar{x}\int_{0}^{\infty} \frac{\diff T}{T}\frac{ \mathrm{e}^{-m^2 T}}{(4 \pi T)^2} \, \frac{q T E }{\sin(qTE) }\, [-1 - 2\cos( 2 q T E) ]\;.
\end{equation}
Apparently, it is a real quantity, but the presence of poles in the $T$-integral signals that this is not the case.
To extract its imaginary part, we go back to Minkowski space via a Wick rotation, using $T \rightarrow iT$, 
$\mathcal{L} \rightarrow -\mathcal{L}$, 
to obtain the Minkowskian effective Lagrangian 
\begin{equation}
 \mathcal{L}[A]= \int_{0}^{\infty} \frac{\diff T}{T}\frac{ \mathrm{e}^{- i m^2 T}}{(4 \pi T)^2} \left(- 3 \frac{ i q T E }{\sin( i q T E) } + 4  (i q T E) \sin( i q T E) \right)\;. 
\end{equation}
For certain values of proper time, the integral develops poles in the $T$-plane, which in turn produce an imaginary part of the Minkowskian effective action. In fact, from
\begin{equation}
  \operatorname{Im}\mathcal{L}[A] = \frac{\mathcal{L}[A] - \mathcal{L}^*[A]}{2i} = \frac{1}{2i} \int_{- \infty}^{+\infty} \frac{\diff T}{T}\frac{ \mathrm{e}^{- i m^2 T}}{(4 \pi T)^2} \left(- 3 \frac{ i q T E }{\sin( i q T E) } + 4  (i q T E) \sin( i q T E) \right)\; , 
\end{equation}
one finds that the contour must be closed in the lower half-plane, and the imaginary part is determined by the residues 
at the poles of the first integrand function, located at\footnote{They correspond to the zero modes of the differential operator $\Delta_{\mu\nu}^{(t)}$ in Minkowski spacetime except for the value $n=0$, which indicates a UV divergence as discussed at the end of the previous section.}
\begin{equation}
 T = -i \frac{\pi n}{q E}\;, \quad 0<n \in \mathbb{N}\; .
\end{equation} 
The final result is
\begin{equation}
 \operatorname{Im}\mathcal{L}[A] = \frac{3}{16 \pi^3}(q E)^2 \, \sum\limits_{n=1}^{\infty} (-1)^{n+1}\frac{\mathrm{e}^{-\frac{m^2\pi n}{q E}}}{n^2} \,\;. 
\end{equation}

In conclusion, the rate for massive spin-1 particle-antiparticle pair production in the presence of a constant electric field per unit of volume and time $\mathcal{P}:=P/\Delta V \Delta \mathcal{T}$ can be written as
\begin{equation}
 \mathcal{P}_{\mathrm{pair}} \approx -\frac{3}{8 \pi^3}(q E)^2\, \operatorname{Li}_{2}\left(-\mathrm{e}^{-\frac{m^2 \pi }{q E}}\right)\;, 
 \end{equation}
where $\operatorname{Li}_2 (\cdot)$ is the polylogarithm of order ${2}$.\footnote{The polylogarithm function is defined by
\begin{equation*}
	\operatorname{Li}_s (z):=\sum\limits_{n=1}^{\infty}\frac{z^n}{n^s}\;.
\end{equation*}}
As already noted in \cite{Vanyashin:1965ple}, this probability corresponds to three times the probability of the production of pairs of scalar particles with mass $m$.

\section{Fermionic spinning particle}\label{sec4}
With the aim of completeness and to compare with our previous method, we present here an alternative first-quantized derivation of the same results. We make use of the worldline model dubbed $\mathcal{N}=2$ massive spinning particle \cite{Bastianelli:2005uy}, which contains fermionic oscillators.
\paragraph{Worldline action} 
The action reads
\begin{equation}
 S= \int \diff \tau \left[ p_\mu \dot{x}^\mu + i \bar{\psi}_\mu \dot{\psi}^\mu + i \bar{\theta} \dot{\theta} - e H - i\bar{\chi} Q - i \chi \bQ- a J_c \right]\;,
\end{equation}
where the main difference with respect to the bosonic theory \eqref{massless bosonic action} consists of the presence of \emph{fermionic} oscillators employed to describe the spin degrees of freedom: their Poisson brackets read
\begin{equation}
 \{\psi^\mu, \bpsi^\nu\}_{\mathrm{PB}} = -i \eta^{\mu\nu}\;, \quad \{\theta, \btheta\}_{\mathrm{PB}} =- i\;,
\end{equation}
and will be translated into anticommutation relations upon quantization. The first-class constraints
\begin{gather}
H = \frac{1}{2}(p^\mu p_\mu + m^2) \;, \quad Q = \psi ^\mu p_\mu +\theta m \;, \quad \bar{Q} = \bar{\psi}^\mu p_\mu +\btheta m\;, \quad J_c = \psi^\mu \bar{\psi}_\mu + \theta \btheta - c\;,
\end{gather}
form the algebra of $\mathcal{N}=2$ supersymmetry in $(0+1)$-dimension, with $J_c$, 
that contains a shift corresponding to the CS coupling $c$, 
acting as the generator of the internal $R$-symmetry:
\begin{gather}
 \{\bar{Q}, Q\}_{\mathrm{PB}} = -2iH \;, \quad \{Q, J_c\}_{\mathrm{PB}} = iQ \;, \quad \{\bar{Q}, J_c\}_{\mathrm{PB}} = -i\bar{Q}\;.
\end{gather}
Under a gauge transformation generated via Poisson brackets by
\begin{equation}
V = \epsilon H + i \bar{\xi} Q + i \xi \bar{Q} + \alpha J_c\,,
\end{equation}
the phase-space variables transform according to
\begin{subequations}
\begin{align}
\delta x^\mu &= \epsilon\, p^\mu + i \xi\, \bar{\psi}^\mu + i \bar{\xi}\, \psi^\mu\;, \\
\delta p_\mu &= 0, \\
\delta \psi^\mu &= -\, \xi\, p^\mu + i \alpha\, \psi^\mu\;, \\
\delta \bar{\psi}^\mu &= -\, \bar{\xi}\, p^\mu - i \alpha\, \bar{\psi}^\mu\;, \\
\delta \theta &= -\, \xi\, m + i \alpha \theta \;, \\
\delta \bar{\theta} &= -\, \bar{\xi}\, m - i \alpha\, \bar{\theta}\;,
\end{align}
\end{subequations}
while the gauge fields 
\begin{subequations}
\begin{align}
\delta e &= \dot{\epsilon} + 2i\, \bar{\chi}\, \xi + 2i\, \chi\, \bar{\xi}\;, \\
 \delta \chi &= \dot{\xi} - i a \xi + i \alpha \chi\;, \\
 \delta \bar{\chi} &= \dot{\bar{\xi}} + i a \bar{\xi} - i \alpha \bar{\chi}\;, \\
 \delta a &= \dot{\alpha}\;.
\end{align}
\end{subequations}
\paragraph{Worldloop path integral and DOF} 
To construct the path integral over the circle and compute the number of degrees of freedom propagated in the loop, we choose antiperiodic boundary conditions (ABC) for the fermionic fields and periodic boundary conditions for the bosonic ones. Similarly to the bosonic case, the gauge symmetries with the chosen boundary conditions allow us to set
\begin{equation}
 G=(e,\bar \chi, \chi, a) \;\to\; \hat{G}=(2T, 0, 0, \phi)\;.
\end{equation}
After Wick rotating, the path integral with Euclidean configuration space action reads
\begin{equation}
 \Gamma = - \int_{0}^{\infty} \frac{\diff T}{T} \mathrm{e}^{-m^2 T} \int\frac{ \diff^D \bar{x}}{(4 \pi T)^{\nicefrac{D}{2}}} \,\mathrm{DoF}(p, D)\;,
\end{equation}
with the number of degrees of freedom given by 
\begin{equation}
 \mathrm{DoF}(p, D) = \int_{0}^{2\pi} \frac{\diff \phi}{2\pi} \,
\mathrm{e}^{ i\left(\frac{1 - D}{2} + p \right) \phi}  
 \left(2\cos \frac{\phi}{2} \right)^{D-1}\;,
\end{equation}
where have set the quantized CS coupling to $c=\frac{1-D}{2} +p$.
The cosines in this expression arise from the integration over the fermionic phase-space variables and from the Faddeev-Popov determinants associated with the SUSY ghosts, which are now bosonic. 
Its calculation leads to
\begin{equation}
 \mathrm{DoF}(p, D) = \binom{D-1}{p}\;.
\end{equation}
It corresponds to the number of degrees of freedom of a massive $p$-form in $D$ spacetime dimensions.
\paragraph{BRST quantization} 
Upon quantization, the Hilbert space $\mathcal{H}_{\mathrm{matter}}$ consists of the states: 
\begin{equation}
 \begin{split}
 \ket{\varphi} &= \sum_{j=0}^{D} (\ket{\varphi_j} + \ket{\varphi^ {(\theta)}_ j}) \\
 & = \sum_{j=0}^{D} \left(\frac{1}{j!} \, \varphi_{\mu_1 ... \mu_j}(x)\, \psi ^{\mu_1} ... \psi ^{\mu_j} \ket{0} + \frac{1}{j!} \, \varphi^{(\theta)}_{\mu_1 ... \mu_j}(x)\, \theta \,\psi ^{\mu_1} ... \psi ^{\mu_j} \ket{0}\right)
 \end{split}
\end{equation}
with $\varphi_{\mu_1 ... \mu_j}(x)$ and $\varphi^{(\theta)}_{\mu_1 ... \mu_j}(x)$ rank-$j$ antisymmetric tensors. Proceeding with BRST quantization along the lines of \cite{Fecit:2023kah} to build a positive definite Hilbert space, one enlarges the phase space with the ghost pairs
\begin{gather}
 \{b, c\} = 1\;, \quad [B,\bar{C}] = 1\;, \quad [\bar{B},C] = 1\;, 
\end{gather}
associated with $(H, Q, \bar{Q})$ respectively. Note that the pairs associated with the SUSY charges are now bosonic. Their ghost number assignments are $\mathrm{gh}(c, \bar{C},C ) = 1$ and $\mathrm{gh}(b, B ,\bar{B}) = -1$. From these operators, the full Hilbert space $\mathcal{H}_{\mathrm{BRST}}$ is then constructed as described in section \ref{sec2} for the bosonic case. The nilpotent BRST charge is
\begin{equation}
 \label{eq:def Q}
 \mathcal{Q}= cH + \bar{C} Q + C \bar{Q} -2C \bar{C} b\;. 
\end{equation}
Once again, the ghost number operator $G = cb + C \bar{B} + B \bar{C} $ and the occupation number operator $\J_p = \psi_\mu \bar{\psi}^\mu + \theta \bar{\theta} + C \bar{B} - B \bar{C} - p$ are introduced.\footnote{With the choice for the CS coupling $c =- \frac{D+1}{2} + p +1$.} They satisfy 
\begin{equation}
 [G, \J_p] = 0\;, \quad [G, \Q] = \Q\;, \quad [\J_p, \Q] = 0\;,
\end{equation}
and are used to define physical states as those states in the cohomology of $\Q$ with vanishing ghost number and occupation number (as measured by $\J_p$).
 The physical states at $p=1$ are contained in the wavefunction

\begin{equation}\label{eq: s=1 state ferm}
 \ket{\psi} = W_\mu(x) \psi ^\mu \ket{0} -i \varphi (x) \theta \ket{0} -f(x) c B \ket{0}\;,
\end{equation}
where, requiring $\ket{\psi}$ to be Grassmann-odd, the Grassmann parities and ghost numbers of the component fields 
$W_\mu(x), \varphi (x), f(x)$ are all vanishing. 
The field equations, obtained by computing  $\Q \ket{\psi}=0$, are found to be the same as the ones reported in \eqref{Proca0}--\eqref{algebraic}.
Upon eliminating the auxiliary field $f$, the gauge symmetries derived from $\delta \ket{\psi}=\Q \ket{\xi}$ correspond exactly to those in Eq.~\eqref{gauge}
Hence, the cohomology at $p=1$ coincides with the one obtained from the bosonic worldline model, with $W_\mu(x)$ being the massive spin-1 field.
\paragraph{Interacting theory} 
The coupling with the background field $A_\mu(x)$ is realized by 
\begin{gather}
 Q \rightarrow \psi^\mu \pi_\mu + \theta m\;, \quad \bQ \rightarrow \bpsi^\mu \pi_\mu + \btheta m\;, \quad H \rightarrow \frac{1}{2}\left(\pi ^2 + m^2 + 2i q F_{\mu \nu} \psi ^\mu \bpsi^\nu \right)\;.
\end{gather}
The squared deformed BRST charge now reads
\begin{equation}
\begin{split}
 \Q_A^2 &= -\frac{i}{2}c \left(\bar C \psi^\rho - C \bpsi^\rho \right) \partial^\mu \tilde{F}_{\mu \rho} 
 -i c \, \bar C  \psi ^\mu \psi^\rho \bpsi ^\nu \partial_\rho \tilde{F}_{\mu \nu} 
 +i c \, C  \psi ^\mu \bpsi ^\nu \bpsi^\rho \partial_\rho \tilde{F}_{\mu \nu} 
 \\
 &\phantom{=}+ \bar C^2  \psi ^\mu \psi ^\nu \tilde{F}_{\mu \nu}  + C^2  \bpsi ^\mu \bpsi ^\nu \tilde{F}_{\mu \nu}  - C \bar C  \psi ^\mu \bpsi ^\nu \tilde{F}_{\mu \nu}\;.
\end{split}
\end{equation}

In general, it is not zero. However, when its action is restricted to the subspace $p=1$, see \eqref{eq: s=1 state ferm}, all but the first term vanish
\begin{equation}
 \Q_A^2\big\vert_{p=1} = -\frac{i}{2}c \left(\bar C \psi^\rho - C \bpsi^\rho \right) \partial^\mu\tilde{F}_{\mu \rho}\;.
\end{equation}
Once again, we find that nilpotency is achieved if the background electromagnetic field $A_\mu$ is on-shell.
\paragraph{Effective action in EM background} 
To construct the path integral for the interacting theory in the $p= 0,1$ sectors, one proceeds just as in the free theory. Taking a constant background electromagnetic field in the Fock-Schwinger gauge \eqref{FS}, the path integral becomes 
\begin{equation}
\begin{split}
 \Gamma[A]&= - \int\diff^4 \bar{x} \int_{0}^{\infty} \frac{\diff T}{T} \mathrm{e}^{-m^2 T} \int_{0}^{2\pi} \frac{\diff \phi}{2\pi} \, \mathrm{e}^{i\left(-\frac{3}{2}+ p\right) \phi}\; \frac14 \cos^{-2}\left(\frac{\phi}{2}\right) \int_{\mathrm{DBC}} Dt \, \mathrm{e}^{-S_t[X,\hat{G}; A]} \\ 
 &\phantom{=\;} \int_{\mathrm{ABC}} D\bpsi D\psi \, \mathrm{e}^{-S_\psi[X,\hat{G}; A]} \int_{\mathrm{ABC}}D\btheta D\theta  \, \mathrm{e}^{-S_\theta[X,\hat{G}]}\;,
\end{split}
\end{equation} 
where
\begin{align}
 S_t[X,\hat{G};A] &= \int \diff\tau \, \frac{1}{2} t^\mu \Delta_{\mu\nu}^{(t)} t^\nu\;, \quad \text{with} \quad \Delta_{\mu\nu}^{(t)}=- \frac{1}{2T} \delta_{\mu \nu}\frac{\diff^2}{\diff\tau^2} - i q F_{\mu \nu}\frac{\diff}{\diff\tau}\;,\\
 S_\psi[X,\hat{G};A] &= \int \diff\tau \, \bar\psi^\mu \Delta_{\mu\nu}^{(\psi)}\psi^\nu\;, \quad \hspace{-0.05cm}\text{with} \quad \Delta_{\mu\nu}^{(\psi)}=\delta_{\mu\nu}\left(\frac{\diff}{\diff\tau} - i \phi \right) + 2i q T F_{\mu \nu}\;,\\
 S_\theta[X,\hat{G}] &= \int \diff\tau \,\btheta \Delta_{\mu\nu}^{(\theta)}\theta\;, \quad \hspace{0.55cm}\text{with} \quad \Delta_{\mu\nu}^{(\theta)} =\frac{\diff}{\diff\tau} - i \phi\;.
\end{align}
Evaluating the functional determinants, we get
\begin{equation}
 \Gamma[A] = - \int \diff^4 \bar{x} \int_{0}^{\infty} \frac{\diff T}{T}\frac{ \mathrm{e}^{-m^2 T}}{(4 \pi T)^2} \,{\rm det}^{-\nicefrac{1}{2}} \left(\frac{\sin(q T F_{\mu \nu})}{q T F_{\mu \nu}}\right)\, I_p(T, A)\;,
\end{equation}
with 
\begin{equation}
 I_p(T, A) = \int_{0}^{2\pi} \frac{\diff \phi}{2 \pi} \, \mathrm{e}^{i\left(-\frac{3}{2}+ p\right) \phi} \;8\cos^{-1}\left(\frac{\phi}{2}\right) \,{\rm det} \left[2\cos\left(-\frac{\phi}{2}\delta_{\mu \nu} + q T F_{\mu \nu} \right)\right] \;. 
\end{equation}
Diagonalizing the field strength $F_{\mu \nu}$, cf. \eqref{diag1}--\eqref{diag2}, the modular integration in the Wilson variable $z = \mathrm{e}^{-i\phi}$ becomes
\begin{equation}\label{I -phi int}
 I_p(T,A) = \oint \frac{\diff z}{2 \pi i} 
\frac{1}{z^{p +1}}  
 \frac{(1 + z^2)^2 + 2z (1 + z^2)(\mathcal{K}_+ + \mathcal{K}_-) + 4z^2 \mathcal{K}_+ \mathcal{K}_-}{(z+1)}\;,
\end{equation}
where $\mathcal{K_{+}}= \cosh( 2 q T K_{+})$ and $\mathcal{K_{-}}= \cos( 2 q T K_{-})$. Deforming the contour to avoid the pole in $z=-1$, we find:
\begin{enumerate}[label=(\roman*)]
 \item For the $p = 0$ case
 \begin{align}
 I_0(T, A) = \mathrm{Res}\left[ \frac{(1 + z^2)^2 + 2z (1 + z^2)(\mathcal{K}_+ + \mathcal{K}_-) + 4z^2 \mathcal{K}_+ \mathcal{K}_-}{z (z+1)} \right]_{z=0} = 1\;.
\end{align}
\item For the massive 1-form case
\begin{align}
 I_1(T, A) = \mathrm{Res}\left[\frac{(1 + z^2)^2 + 2z (1 + z^2)(\mathcal{K}_+ + \mathcal{K}_-) + 4z^2 \mathcal{K}_+ \mathcal{K}_-}{z^2 (z+1)}\right]_{z=0} =-1+ 2(\mathcal{K}_+ + \mathcal{K}_-)\;.
 \end{align}
 \end{enumerate}
 Thus, we have reobtained the same results found in previous sections. In particular, setting $p=0$ we get the usual Weisskopf effective Lagrangian, whereas for $p=1$ we obtain \eqref{effect action}.

\section{Effective interactions on the worldline}\label{sec5}
The particle models that we have been considering so far can be further extended to include non-minimal couplings to electromagnetism, 
readily interpreted as effective interactions. To illustrate this, let us consider the case of a scalar particle,
which does not require additional oscillators in its simplest formulation.
The action in phase space is given by
\begin{equation} 
S[x,p,e]=\int \diff\tau \left( p_\mu \dot{x}^\mu -eH  \right)\;, 
\end{equation}
where the Hamiltonian constraint $H$ has the form
\be
H= \frac12 (\pi^\mu \pi_\mu + m^2)\;, \quad \pi_\mu= p_\mu - q A_\mu(x)\;.
\ee
This expression includes the minimal coupling to the abelian gauge background $A_\mu(x)$, 
implemented by the covariant momentum $\pi_\mu$.
It corresponds to the Klein-Gordon operator upon quantization. 

To model additional, gauge-invariant, non-minimal couplings to $A_\mu$, we deform the Hamiltonian as follows
\be
H \quad \to \quad H'= \frac12 (\pi^\mu \pi_\mu + m^2)  + c_1 F_{\mu\nu} F^{\mu\nu} + c_2 \partial_\lambda 
F_{\mu\nu}  \partial^\lambda F^{\mu\nu} +\cdots \;.
\ee
Here, $c_1$, $c_2$, etc.,  are effective couplings with negative mass dimensions,
corresponding to non-renormalizable interactions in standard quantum field theory. These terms are expected to encode the 
consequences of the particle's internal structure.
The study of analogous terms in particle and string theory has a long history, exemplified by the L\"uscher term of the QCD string \cite{Luscher:1980fr, Luscher:1980ac}, the worldline curvature contributions discussed in \cite{Pisarski:1986gp, Awada:1993bv, Baig:1996se}, and the more recent Goldberger–Rothstein approach to classical black hole scattering \cite{Goldberger:2004jt, Goldberger:2007hy, Porto:2016pyg}.

The worldline formalism can be employed to study the implications of these effective couplings, particularly on the rate of pair production.
To make this explicit, let us consider the simplest case and focus on the $c_1$ coupling with constant $F_{\mu\nu}$. 
In this case, the term $c_1 F_{\mu\nu} F^{\mu\nu}$ can be produced by the shift 
\be
m^2 \quad \to  \quad m^2 + 2 c_1 F_{\mu\nu} F^{\mu\nu} 
\ee
acting on the original action.
For a constant electric field of magnitude $E$, the shift reduces to $m^2 \to m^2 - 4 c_1 E^2$.
Therefore,  from the well-known leading term of the pair production rate for the scalar particle 
\begin{equation}
 {\cal{P}_{\text{pair}}}  = \frac{(q E)^2}{8 \pi^3} \mathrm{e}^{-\frac{\pi m^2}{q E}}\;,
\end{equation}
we immediately obtain the modified expression which includes the effective coupling  $c_1$
\be
 {\cal{P}_{\text{pair}}}  = \frac{(q E)^2}{8 \pi^3} \mathrm{e}^{-\frac{\pi m^2}{q E} + \frac{4\pi c_1}{q} E}  \;.
\ee
Naturally, the value for $c_1$ must either be determined experimentally or derived from a more refined theoretical model describing the extended nature of the particle. 
A naive expectation is that pair production should be suppressed
as, intuitively, we believe more difficult to create a composite 
object like a house rather than a more elementary object like a brick,
suggesting that admissible values of $c_1$
are likely negative (note that $E=\sqrt{\vec{E}^2}>0$, while $q=|q|$ in these formulae).
Nevertheless, more refined analysis might lead to the opposite conclusion, as exemplified 
in \cite{Lebedev:1984mei}, where a two-loop correction to the Euler-Heisenberg Lagrangian was seen to give rise to an enhancement of the electron/positron pair creation.
However, a detailed investigation of this question lies beyond the scope of this work.

A similar correction applies to the pair production rate for the massive spin-1 particle.
Worldline instanton methods could now be applied to treat more general scenarios, including non-constant electromagnetic backgrounds and additional effective couplings.

\section{Conclusions} \label{sec6}
In this paper, we have constructed worldline actions for a massive, charged, spin-1 particle. These actions carry the necessary gauge symmetries to ensure a unitary description at the quantum level. We presented two equivalent formulations: one employing bosonic oscillators and the other using fermionic oscillators on the worldline. A crucial feature in both formulations is the gauging of the oscillator number operator, appropriately shifted by a Chern-Simons coupling, which enforces the projection onto the spin-1 sector.
A BRST analysis reveals that the spin-1 sector can couple to an external electromagnetic field in both formulations, provided the field satisfies the vacuum Maxwell equations. This condition suffices for applications such as computing scattering amplitudes involving external photons, whose asymptotic states obey these equations. We employed this coupling to analyze the system's behavior in a constant electromagnetic field.
The spin-0 sector -- also present in both models -- admits coupling to the electromagnetic field without requiring any conditions on the field. In contrast, other sectors corresponding to higher-spin particles (in the bosonic model) and antisymmetric tensor fields of rank 
$p>1$ (gauge  $p$-forms) do not admit electromagnetic couplings within our worldline framework.

By studying the path integral of the spin-1 particle action on the circle, we obtained the one-loop effective action. For a constant electromagnetic background, we explicitly computed the corresponding effective Lagrangian of the Euler-Heisenberg type. From this, we derived the pair production rate for spin-1 particle-antiparticle pairs in a constant electric field. Our results fully agree with the quantum field theory result originally presented in \cite{Vanyashin:1965ple}, which we recover here using a purely worldline approach, independent of any second-quantized formalism.
Additional, more recent works, discussing the effective Lagrangian induced by spin-1 particles and related matters, include 
\cite{Batalin:1976uv, Skalozub:1976tr, Dittrich:1983ej, Reuter:1984zk, Blau:1988iz,         
Jikia:1993tc, Preucil:2017wen, Henriksson:2021ymi, Alviani:2024sxx}.

Alternative worldline formulations have been explored in the literature, notably in \cite{Reuter:1996zm}, where worldline techniques are employed to express the heat kernel representation of the spin-1 QFT effective action as worldline path integrals, allowing for a more efficient computation, including the evaluation of worldline functional determinants.

Finally, we have explored extensions of our worldline models to include additional effective interactions with the electromagnetic field, which may impact the pair production rate. In the simple case of a constant electric field, the resulting modifications are captured by a shift in the particle's mass within the known pair production formulas. More general field configurations require a dedicated analysis which may benefit of the worldline instanton techniques \cite{Affleck:1981bma, Dunne:2005sx}.

Our results add to the effort of constructing wordline methods without using key input from QFT, 
following the path employed in the study of Yang-Mills theory \cite{Dai:2008bh, Bastianelli:2025xx}, gravity \cite{Bonezzi:2018box, Bonezzi:2020jjq, Bastianelli:2019xhi, Bastianelli:2022pqq, Bastianelli:2023oca, Fecit:2023kah, Fecit:2024jcv}, and scalar theories \cite{Bonezzi:2025iza}.

\section*{Acknowledgments}
We thank the organizers of the ``First Quantisation for Physics in Strong Fields" workshop (Higgs Centre for Theoretical Physics, University of Edinburgh) for fostering a stimulating environment, and we further thank the participants for insightful discussions, both of which contributed to the development of this work. We benefited from valuable suggestions from conversations with Roberto Bonezzi, Olindo Corradini, Sebastián~A.~Franchino-Vi\~nas, Markus Fr\"ob, Arttu Rajantie, Davide Rovere and Christian Schubert.

\appendix

\section{Functional determinant from the Gel'fand--Yaglom theorem}\label{appA}
In this section, we detail the computation of the functional determinant \eqref{Det1} associated with the differential operator
\begin{align}\label{op}
 \Delta_{\mu\nu}^{(t)}(\tau,\tau')&=\left [
 - \frac{1}{2T} \delta_{\mu \nu}\frac{\diff^2}{\diff\tau^2}  - i q F_{\mu \nu}\frac{\diff}{\diff\tau}\right ]\delta(\tau-\tau')\;,
\end{align}
using (generalizations of) the Gel'fand--Yaglom theorem \cite{Gelfand:1959nq}. In its simplest formulation, the GY theorem states that given a one-dimensional second-order differential operator defined on the interval $z \in [0,1]$ and the associated eigenvalue problem with vanishing Dirichlet boundary conditions
\begin{equation} \label{Laplace}
\left[ -\frac{{\rm d}^2}{{\rm d} z^2}+V(z) \right] \psi(z)=\lambda \, \psi(z)\ , \quad \text{with} \quad \psi(0)=\psi(1)=0\;,
\end{equation}
if we can solve the initial value problem
\begin{equation}
\left[ -\frac{{\rm d}^2}{{\rm d}z^2}+V(z) \right] \Phi(z)=0\ , \quad \text{with} \quad \Phi(0)=0\ , \quad \dot{\Phi}(0)=1\;,
\end{equation}
then the boundary value of the solution determines the functional determinant of the differential operator
\begin{equation} \label{GY1D}
\mathrm{Det} \left[ -\frac{{\rm d}^2}{{\rm d}z^2}+V(z) \right] \propto \Phi(1)\;.
\end{equation}
The precise definition of the functional determinant actually involves the ratio of two determinants, and should be understood in this sense \cite{Dunne:2007rt}; in this work, we compute the determinants relative to the corresponding determinant for the free operators upon extracting the zero modes of the latter. For our purposes, we need the generalization of the theorem for higher-dimensional operators: we report the main formulae, while referring the reader to the references \cite{Kirsten:2003py, Kirsten:2004qv, Fecit:2025kqb} for more details. \\

To compute the functional determinant of $\Delta_{\mu\nu}^{(t)}$ acting on the quantum fluctuations $t^\mu(\tau)$ with Dirichlet boundary conditions, one needs to solve the associated homogeneous differential equation for the solution with initial conditions
\begin{equation}\label{initDBC}
\varphi_\mu^{(\rho)}(0)=0\;, \quad \dot{\varphi}_\mu^{(\rho)}(0)=\delta^\rho_\mu\;,
\end{equation}
and then 
\begin{equation}
 \mathrm{Det}\left(\Delta_{\mu\nu}^{(t)}\right)=\mathrm{det}\left[\varphi_\mu^{(\rho)}(1) \right]\;.
\end{equation} 
Using a simple trick, we can recast the operator \eqref{op} into a form suitable for a direct application of the theorem. In particular, it is not hard to show that
\begin{equation}
 \Delta_{\mu\nu}^{(t)}(\tau,\tau') \propto \left[ \mathrm{e}^{-iq F T \tau} \left(-\frac12\frac{\diff^2}{\diff\tau^2}\delta(\tau-\tau') -\frac12 q^2 F^2\delta(\tau-\tau') \right) \mathrm{e}^{i q FT \tau} \right]_{\mu\nu}\;,
\end{equation}
from which one sees that its functional determinant can be equivalently computed by calculating the determinant of the operator inside round brackets.\footnote{The prefactors hidden in ``$\propto$" are taken care of by normalizing with the corresponding free operator.} Solving for the associated homogeneous equation with initial conditions \eqref{initDBC} we get
\begin{equation}
 \varphi_\mu^{(\rho)}(z)=\left[ \frac{\sin(qFTz)}{q FT} \right]^\rho_\mu\;,
\end{equation}
which confirms Eq.~\eqref{Det1}.

\addcontentsline{toc}{section}{References}
\printbibliography

\end{document}